\documentclass[12pt,preprint]{aastex}
\usepackage{graphicx}
\usepackage{apjfonts}
\usepackage{natbib}

\begin{document}

\title{On the nature of the first transient Z-source XTE~J1701$-$462: its accretion disk structure, neutron star magnetic field strength, and hard tail}

\author{G. Q. Ding\altaffilmark{1}, S. N. Zhang\altaffilmark{2}, N. Wang\altaffilmark{1}, J. L. Qu\altaffilmark{2}, and S. P. Yan\altaffilmark{1,3}}
\altaffiltext{1}{Xinjiang Astronomical Observatory, Chinese Academy of Sciences, 150, Science 1-Street, Urumqi, Xinjiang 830011, China; dinggq@uao.ac.cn, dinggq@gmail.com}
\altaffiltext{2}{Key Laboratory of Particle Astrophysics, Institute of High Energy Physics, Chinese Academy of Sciences, Beijing 100049, China}
\altaffiltext{3}{Graduate School of Chinese Academy of Sciences, Beijing 100049, China}

\begin{abstract}

Using the data from the Rossi X-Ray Timing Explorer satellite, we
investigate the spectral evolution along a ``Z'' track and a ``$\nu$''
track on the hardness-intensity diagrams of the first transient Z
source XTE~J1701$-$462. The spectral analyses suggest that the inner
disk radius depends on the mass accretion rate, in agreement with the
model prediction, $R_{\rm in} \propto {\dot{M}_{\rm disk}}^{2/7}$, for
a radiation pressure dominated accretion disk interacting with the
magnetosphere of a neutron star (NS). The changes in the disk mass
accretion rate $\dot{M}_{\rm disk}$ are responsible for the evolution
of the ``Z'' or ``$\nu$'' track. The radiation pressure thickens the
disk considerably, and also produces significant outflows. The NS
surface magnetic field strength, derived from the interaction between
the magnetosphere and the radiation pressure dominated accretion disk,
is $\sim$(1--3)$\times10^9$~G, which is possibly between normal atoll
and Z sources. A significant hard tail is detected in the horizontal
branches and we discuss several possible origins of the hard tail.

\end{abstract}


\section{INTRODUCTION}

According to their timing properties and shapes of X-ray color-color
diagrams (CDs) or hardness-intensity diagrams (HIDs), low mass neutron star
X-ray binaries (NSXBs) can be divided into two classes: Z sources and
atoll sources (Hasinger \& van der Klis 1989). Six Galactic NSXBs have been
classified as Z sources: Sco~X-1, GX~17+2, GX~349+2, GX~5$-$1, GX~340+0, and
Cyg~X-2. As shown in Figure 1a of the paper of Hasinger \& van der Klis (1989),
the tracks in the CDs of Cyg~X-2, GX~5$-$1, and GX~340+0 (the so-called Cyg-like
Z sources) resemble the letter ``Z''. However, the tracks in the CDs of Sco~X-1,
GX~17+2, and GX~349+2 (the so-called Sco-like Z sources) constitute a pattern
looking like the Greek letter ``$\nu$''. From top to bottom, the tracks in the
CDs of Z sources are divided into three branches which are called horizontal
branch (HB), normal branch (NB), and flaring branch (FB). The CDs of atoll
sources are different from those of Z sources. Generally, a complete CD of an
atoll source consists of three parts which are called island state (IS), lower
banana (LB), and upper banana (UB), respectively. However, an additional
part, i.e., an extreme island state (EIS) is present in some
cases (Homan et al. 2010).

In addition to the six Z sources, other two NSXBs also show the
characteristics of Z sources in most cases. One is Circinus~X-1 (Cir X-1).
Shirey, Bradt, \& Levine (1999) studied the timing and spectral behaviors
of Cir X-1 and considered this source as a Z source. Ding, Qu, \& Li (2003)
investigated the evolution of the hard X-ray spectra along the ``Z'' track of
Cir~X-1 and found that the evolution behaviors are similar to those of other Z
sources. However, it was reported that Cir~X-1 also exhibited the behaviors
of atoll sources at lower intensity (Oosterbroek et al. 1995). The other is
XTE~J1701$-$462, the first new Z source in nearly 35 years, which was first
detected with the All-Sky Monitor (ASM) on board the Rossi X-Ray Timing
Explorer (RXTE) on January 18, 2006 (Remillard \& Lin 2006). Homan et al.
(2006a, 2006b, 2007a) suggested that XTE~J1701$-$462 was a Z source, because it
exhibited typical CDs of Z sources, and its timing properties were also
consistent with those of Z sources. However, it was also reported that
XTE~J1701$-$462 evolved into an atoll source at the end of its outburst
(Homan et al. 2007b). As shown in Figure 4 in the paper of Homan et al. (2007a),
XTE~J1701$-$462 showed two types of tracks on the CDs: the ``Z'' tracks and
the ``$\nu$'' tracks during the Z-source stages.

Lin, Remillard, \& Homan (2009b) analyzed the RXTE observations of
XTE~J1701$-$462 during the 2006-2007 outburst and concluded that during the
Atoll-source stage the disk or neutron star (NS) evolved roughly along
the $L_{\rm X}\propto T^4$ track, both the inner disk radius and blackbody
radius of the NS remained constant, and the inner disk radius, comparable with
the NS radius, was set by the innermost stable circular orbit (ISCO). However,
during the Z-source stages the disk or NS departed from the $L_{\rm X}\propto T^4$
track and the inner disk radius increased with the source intensity, which could
be a signature of the local Eddington limit in the disk. These results lead us to
the thought that the magnetosphere of the NS could play a crucial role in the
evolution of inner disk radius and disk structure. It is plausible that the inner
edge of the disk cannot go across the the ISCO or magnetospheric radius. During
the Atoll-source stage, the gas pressure from the disk pushes the magnetosphere
to near the ISCO, so the inner disk radius is set by the ISCO. However, during
the Z-source stages, because the disk is thickened by the radiation pressure, the
gas pressure from the disk decreases, magnetosphere expands, and then the inner
disk radius increases. Therefore, during the Z-source stages the inner disk
radius could be set by the magnetospheric radius and it should vary with the
mass accretion rate ($\dot{M}$).

Hasinger \& van der Klis (1989) considered two scenarios for the relation 
between Z and atoll sources: (i) the two types of sources are similar and their 
differences are due to the difference of their $\dot{M}$; (ii) their differences 
are resulted from some additional parameters such as neutron star surface magnetic 
field strength, disk inclination, etc., rather than $\dot{M}$. They argued that 
under scenario (i), one could expect that in the CD the atoll and Z patterns would 
evolve into each other in the sources with a large range in $\dot{M}$ and these 
sources then switch between Z and atoll behaviors. Because this kind of evolution 
had never been observed, they excluded this scenario. Therefore, they proposed 
scenario (ii) is responsible for the difference between atoll sources and Z sources 
and suggested that both the mass accretion rate and neutron star surface magnetic 
field strength might be larger in Z sources than in atoll sources, which might be 
due to different evolutionary history. However, it was observed that during the 
2006-2007 outburst of XTE~J1701$-$462, the track in CD evolved from the ``Z'' track, 
via the ``$\nu$'' track, and finally to the atoll pattern (Lin, Remillard, \& Homan 
2009b; Homan et al. 2010). This finding thus strongly supports the first scenario 
considered by Hasinger \& van der Klis (1989).

XTE~J1701$-$462 thus plays a role in linking between Z and atoll sources (Lin, Remillard, 
\& Homan 2009b). In order to understand this critical source more clearly and know 
how a Z source evolves into an atoll source, its characteristics, including its mass 
accretion rate, the structure and evolution of the accretion disk, NS magnetic field, 
and hard X-ray emission are needed to be studied. The normal Z sources exhibit 
complicated behaviors of hard tail (D'Amico et al. 2001; Di Salvo et al. 2000, 2001, 
2002, 2006; Ding et al. 2003, 2006; Iaria et al. 2001, 2002), but so far the hard tail 
behaviors of XTE~J1701$-$462 are unclear so far. The NS surface magnetic field strength 
is believed to be higher in Z sources than in atoll sources (Focke 1996; Zhang, Yu, \&
Zhang 1998), whereas it is still unknown in XTE~J1701$-$462. During the 2006-2007 
outburst of XTE~J1701$-$462, with dramatic change of its intensity, it evolved from a 
Cyg-like Z source, via a Sco-like Z source, to an atoll source, so the evolution was 
likely linked to its mass accretion rate, rather than its magnetic field strength, 
because it was unlikely that its magnetic field strength varied in such a short term. 
Since this transient Z source evolves into an atoll source at low luminosity, it is 
natural to speculate that its magnetic field strength keeps unvaried during its 
evolution and it may be between the two types of NSXBs, although there is no obvious 
dichotomy between the magnetic field strengths in atoll and Z sources, as implied by 
the scenario (i) of Hasinger \& van der Klis (1989). It is then understandable that at 
low or high accretion rates the source should appear as an atoll or a Z source, if the 
above speculation is true.

In this work, using the data from the Proportional Counter Array (PCA) and the 
High-Energy X-ray Timing Experiment (HEXTE) on board RXTE satellite, we investigate 
the spectral evolution along a ``Z'' track and a ``$\nu$'' track on its HIDs of the 
first transient Z source XTE~J1701$-$462, study its accretion disk structure and 
evolution, infer its NS surface magnetic field strength, and detect the anticipated 
hard tail. We describe our data analyses in \S 2, represent the methods for us to 
estimate the NS magnetic field strength and mass accretion rates in \S 3, discuss 
our results in \S 4, and present our conclusions in \S 5.

\section{DATA ANALYSIS}

\subsection{Producing CDs and HIDs}

We select two groups of observations for XTE~J1701$-$462 to perform our
analyses. The observations of the first group were made between 2006 January
21 and January 31 (MJD 53,756--53,766) and those of the second group were
made between 2006 March 2 and March 7 (MJD 53,796--53,801). The selected
observations are listed in Tables 1 and 2. In order to produce CDs and HIDs,
following Homan et al. (2007a), we define the soft color as the ratio of
count rates between the [4.0--7.3~keV (channels 7--14)] and [2.4--4.0~keV
(channels 3--6)] energy bands, and the hard color as that between the
[9.8--18.2~keV (channels 21--40)] and [7.3--9.8~keV (channels 15--20)] energy
bands, and intensity as the count rate covering the energy range
$\sim$2.4--18.2~keV (channels 3--40). The colors and intensity are produced
from the background-subtracted light curves with a time resolution of 16~s,
which are produced from the ``Standard 2'' mode data from proportional counter
unit (PCU) number 2 (the most reliable of the five PCUs). As shown in Figure 1,
the track in CD1, produced from the first group observations, is divided
into five branches which are, respectively, called as upturn HB (UHB), HB, 
NB, FB, and low FB (LFB). However, CD2, produced from the
second group observations, is only split into three branches, i.e., HB, NB,
and FB. These branches are converted into the corresponding branches in HID1
and HID2, which are shown in Figure 2. As shown in Figure 2, the tracks in
HID1 and HID2 are, respectively, divided into eighteen regions and eleven
regions, in order to minimize count rate variations within each of the 29
regions and produce the PCA and HEXTE spectra of each region. In HID1, UHB,
HB, NB, FB, and LFB are composed of regions 1--3, regions 4--9, regions 10--14,
regions 15--16, and regions 17--18, respectively; in HID2, regions 1--2,
regions 3--4, and 5--11 separately belong to HB, NB, and FB.

\subsection{Extracting the spectra}

The ``Standard 2'' mode data from PCU2 are used to extract the PCA spectra.
We use the ``Standard'' mode HEXTE data to produce the HEXTE spectra. HEXTE
comprises two clusters, cluster A and cluster B, each of which is composed
of four detectors. Generally, we use the data from cluster A to produce the
HEXTE spectra, because detector 2 of cluster B loses its spectral capability
and automatic gain control. However, in some cases, the background files
cannot be produced from the data of cluster A, so we have to use the data
from cluster B to produce the HEXTE spectra, and in these cases, we discard
the data from detector 2 of this cluster. In order to have sufficient
statistics for the spectra, especially for the HEXTE spectra, all observation
data within a region in the HIDs are combined together. In order to improve
the quality of the HEXTE spectra, the HEXTE spectra are rebinned with the
criterion: the minimum signal-noise ratio (SNR) of per new bin is 2.5.
Following the recipes for RXTE data analysis, the PCA and HEXTE background
spectra are respectively produced, and then we get the background-subtracted
PCA+HEXTE spectrum of each of the 29 HID regions.

\subsection{Selection of the spectral model}

Using RXTE observations for atoll source Aql~X-1 and 4U~1608-52, Lin,
Remillard, \& Homan (2007b) tested some commonly used NSXB spectral models
such as BPL+MCD (BPL: broken power law, bknpower in XSPEC; MCD: multicolor
disk, diskbb in XSPEC), CompTT+MCD (CompTT: thermal Comptonization, compTT in XSPEC),
CompTT+BB (BB: single-temperature blackbody, bbodyrad in XSPEC), etc., and
evaluated their performance against desirability criteria,
including $L_{\rm X}\propto T^4$ for both the MCD and BB included in these models.
They found that none of the classical models for thermal emission plus Comptonization
performed well in the soft state (SS). Instead, they devised a hybrid model: BB+BPL for
the hard state (HS) and MCD+BB+CBPL (CBPL: constrained BPL) for the SS
and found that in this model both MCD and BB produced $L_{\rm X}\propto T^4$ tracks
well and the BB emission area was roughly constant over a wide range of $L_{\rm X}$.
In this model, the MCD is interpreted as the emission from the accretion disk and it
provides the inner disk radius $(R_{\rm in})$ and the temperature at the inner disk
radius $(T_{\rm in})$; the BB is used to describe the emission from the NS surface
or the boundary layer between the inner disk and the NS, and it provides the BB
temperature $(T_{\rm bb})$ and BB radius $(R_{\rm bb})$. The BPL in the HS
model represents the strong Comptonization emission from a heavily Comptonized disk,
whereas the CBPL in the soft state model accounts for the weak Comptonization
emission from a very weakly Comptonized disk.

Applying their hybrid model developed for observations of atoll sources, Lin,
Remillard, \& Homan (2009b) analyzed all the spectra of 866 observations during
2006-2007 outburst of XTE~J1701$-$462 and investigated how this source evolved
from a Cyg-like Z source, via a Sco-like Z source, and then to an atoll source.
They divided the outburst into five stages. In stage I and II--IV, the source
showed the characteristics of Cyg-like and Sco-like Z sources, respectively, whereas
it showed characteristics of an atoll source in stage V which included the SS and HS.
Their analyses suggested that during the Z-source stages (stages I--IV), the CBPL
component contributed significantly when the source entered the UHB or HB, and the
inclusion of such a component could not improve the fitting significantly for the
spectra of the HB/NB vertex, NB, NB/FB vertex, and FB. In our analyses, the observations
of the first and second groups are within stage I (the Cyg-like Z stage) and stage II
(a Sco-like Z stage), respectively. Following the models used for the Z-source stages,
which are showed in Figure 13 in the paper of Lin, Remillard, \& Homan (2009b), we
use the MCD+BB+CBPL (CBPL: $E_{\rm b}= 20$~keV, ${\Gamma}_1= 2.5$) model to fit the
spectra of regions 1--9 in HID1 (i.e., the UHB and HB spectra in HID1) and the spectra
of regions 1--2 in HID2 (i.e., the HB spectra in HID2), with the CBPL component in this
model to account for the hard tail. Nevertheless, we use the MCD+BB model to fit the
spectra of other regions.

Farinelli, Titarchuk, \& Frontera (2007) proposed that weakening (or
disappearance) of the hard X-ray tail could be explained by the increasing
radiation pressure originated at the surface of the NS and used a model
consisting of a bulk-motion Comptonization (BMC) plus a CompTT to fit the
broadband spectra of GX~17+2. In order to trace the origin of the hard
X-ray emission in XTE~J1701$-$462, we also use this model to fit the
spectra of regions 1--9 in HID1 and the spectra of regions 1--2 in HID2.

\subsection{Spectral fitting}

We use XSPEC 12.5 to jointly fit the PCA and HEXTE spectra of each of the
29 regions. Due to the calibration uncertainties, a systematic error of
1\% is added to the PCA spectra, but not to the HEXTE spectra. The PCA
spectra are taken in 3--21 keV. In order to improve the SNRs of the HEXTE
channels, we rebin the 64 original HEXTE channels and then take the new
bins which have SNRs larger than 2.5 and ignore those bins with SNRs
smaller than 2.5. The PCA and HEXTE response matrices provided by RXTE
team are applied. A multiplicative constant representing the relative
normalization between the instruments is added to the spectral model, for 
taking into account the uncertainties in the PCA-HEXTE cross calibration.

A Gaussian component, accounting for an iron emission line, is also
included in our spectral models and the interstellar absorption is taken
into account. Following Lin, Remillard, \& Homan (2009b), the Gaussian width
$(\sigma)$ of the line component and hydrogen column density $(N_{\rm H})$
of the interstellar absorption are fixed at 0.3 keV
and $2.0\times10^{22}$~cm$^{-1}$, respectively. Lin, Remillard, \& Homan (2009b)
estimated that the orbital inclination is neither very high ($\lesssim75\degr$),
due to absence of eclipses or absorption dips, nor very low because of the
observed narrow equivalent width ($\lesssim$~50 ev) of the iron line. They thus
assumed a binary inclination of $70\degr$ of XTE~J1701$-$462. In our analyses,
we also take the inclination angle to be $70\degr$. From the observed photospheric
radius expansion, Lin et al (2007a, 2009a) estimated the distance to the source to
be 8.8 kpc, which is also used in our analyses.

The fitting results are listed in Tables 3 and 4, and some unfolded spectra
are shown in Figures 3--5. Figures 6 and 7 demonstrate some folded spectra and the
corresponding residual distribution. Figure 8 shows the evolution of the parameters
along the ``Z'' and ``$\nu$'' tracks. We calculate the ratio of the CBPL flux
to the total flux, and the fraction of CBPL is exhibited by Figure 9. In Figure 10,
the luminosities of thermal components (MCD/BB) vs. their characteristic
temperatures as well as the PCA intensity are demonstrated, and the characteristic
emission sizes of the thermal components vs. the PCA intensity are also demonstrated.
Figure 11 shows how the source evolves along the ``Z'' and ``$\nu$'' tracks in the
maps of the individual luminosity vs. total luminosity.

\section{Radiation pressure dominated disk interacting with NS's magnetosphere}

Lin, Remillard, \& Homan (2009b) have shown that when the source is in the
Z-source stages, $R_{\rm in}$ starts to increase with the total X-ray luminosity.
They interpreted this as due to the local Eddington luminosity of the disk emission
is reached, i.e., $R_{\rm in}$ is determined by the local Eddington luminosity; this
model would naturally predict $R_{\rm in}\propto L_{\rm x,\ Edd}$. However, as shown
in Figure 19 of the paper of Lin, Remillard, \& Homan (2009b), $R_{\rm in}$ increased
only by a factor of 1.5 when $L_{\rm x,\ Edd}$ increased from 0.4 to 0.9, clearly
deviating $R_{\rm in}\propto L_{\rm x,\ Edd}$. This is understandable because the NS
surface magnetic field strength and mass accretion rate ($\dot{M}$) are two important
parameters of an accreting NS binary system; however the model
of Lin, Remillard, \& Homan (2009b) did not take into account the interaction of the
NS's magnetosphere with the accretion disk.

In this work we attempt to derive the surface magnetic field strength of the
NS from the interaction between the magnetosphere and the accretion disk. It is
reasonable to assume that the inner edge of the accretion disk cannot go across
the surface of the magnetosphere or the ISCO. The magnetosphere will normally 
expand when $\dot{M}$ decreases, for a given accretion disk structure. When the 
magnetospheric radius ($R_{\rm m}$) is greater than 
the ISCO ($3R_{\rm s}=6GM_{\rm ns}/c^{2}$, where $M_{\rm ns}$ is the NS mass),
the inner disk radius ($R_{\rm in}$) is equal to the magnetospheric 
radius, i.e., $R_{\rm in}=R_{\rm m}$ (Zhang, Yu, \& Zhang 1998). This scenario 
is shown in Figure 12. Combining the equation (2) of Cui (1997) and 
equation (18) of Lamb, Pethick, \& Pines (1973), we get
\begin{equation}
R_{\rm in} = 5\left(\frac{B_{\rm a}}{10^{8}\mbox{ }{\rm G}}\right)^{4/7}  \left(\frac{\dot{M}_{\rm
disk}}{\dot{M}_{\rm Edd}}\right)^{-2/7} \left(\frac{M_{\rm ns}}{1.4\mbox{ }M_{\odot}}\right)^{1/7}
\left(\frac{R_{\rm ns}}{10^6\mbox{ }{\rm cm}}\right)^{10/7}\mbox{ }{\rm km},
\end{equation}
where $B_{\rm a}=B_0\sqrt{\xi v_{\rm r}/v_{\rm ff}}$, $B_0$ is the surface
magnetic field strength of the NS, $\xi R_{\rm in}$ is the thickness of the disk
at $R_{\rm in}$, $v_{\rm r}$ and $v_{\rm ff}$ are the radial and free-fall
velocities at $R_{\rm in}$, respectively, while $\dot{M}_{\rm Edd}$ and $R_{\rm ns}$ are
separately the Eddington accretion rate of XTE~J1701$-$462 and the radius of the NS.
The inner disk radius $R_{\rm in}$ and
disk mass accretion rate $\dot{M}_{\rm disk}$ can be inferred from the MCD+BB+CBPL+LINE or
MCD+BB+LINE model, with the known NS mass and radius ($M_{\rm ns}\ {\rm {and}}\ R_{\rm ns}$),
so the NS magnetic field strength can be estimated from above equation.
Because $\xi$, $v_{\rm r}$ and $v_{\rm ff}$ all depend upon the structure of the
disk, so only $B_{\rm a}$ can be inferred directly from observations, without knowing
the details of the disk structure {\it a prior}. For convenience, we
call $B_{\rm a}$ the apparent magnetic field strength of the NS.

In a Keplerian disk, 50\% of the available gravitational potential energy is
converted into radiation at the inner disk boundary, thus the disk mass accretion
rate is computed by the following equation,
\begin{equation}
\dot{M}_{\rm disk} = 2L_{\rm disk}R_{\rm in}/GM_{\rm ns},
\end{equation}
where $G$ is the gravitational constant. $R_{\rm in}$ can be determined
from $L_{\rm disk}$ and inner disk temperature $T_{\rm in}$ by
\begin{equation}
L_{\rm disk} = 4\pi R_{\rm in}^2\sigma (T_{\rm in}/f)^4,
\end{equation}
where $\sigma$ is the Stefan's constant and $f\approx 1.7$ is the color correction
factor (Zhang, Cui, \& Chen 1997). Combining the above two equations, we can then have,
\begin{equation}
\dot{M}_{\rm disk} = f^2L_{\rm disk}^{3/2}/GM_{\rm ns}\sqrt{\pi\sigma}T_{\rm in}^2.
\end{equation}
With the above equation, the best fitting value and the error of $\dot{M}_{\rm disk}$
can be calculated directly from fitted parameters in the spectral model.

Similarly, because almost all gravitational energy is converted into radiation at the
surface of the NS, we can compute the mass accretion rate onto the NS using
equation,
\begin{equation}
\dot{M}_{\rm ns} = L_{\rm ns}R_{\rm ns}/GM_{\rm ns}=3.6\times 10^{-27}\eta L_{\rm ns}/G,
\end{equation}
where $L_{\rm ns}$ is the BB luminosity derived from the spectral
models, $\eta=(R_{\rm ns}/10^6\mbox{ }{\rm cm})(1.4M_{\odot}/M_{\rm ns})$, which is
related to the NS equation of state. Recent measurements of the masses and radii of
several NSs gave $\eta=$0.8--1.2 ({\"O}zel 2006; {\"O}zel et al. 2009; G{\"u}ver et al. 2010a, 2010b).
We therefore take $M_{\rm ns}=1.4$~$M_{\odot}$ and $R_{\rm ns}=10^6$~cm throughout this
paper, unless specified otherwise; this gives $\eta=1$.

When $L_{\rm disk}$ and $L_{\rm ns}$ are computed, a distance of 8.8~kpc to the
source is taken (Lin et al. 2007a, 2009a) and the fluxes in the whole energy band
are used. The fluxes below $\sim$2 keV cannot be estimated from the PCA response 
matrix. However, the disk or NS flux below $\sim$2 keV contributes a considerable 
portion in the whole energy band, because the inner disk temperature ($kT_{\rm in}$) 
or BB temperature ($kT_{\rm bb}$) is lower than several keV. We therefore use the 
XSPEC command ``dummyrsp'' to create a dummy response matrix below $\sim$2 keV and 
thus estimate the fluxes in the whole energy range.

In equation (5), if the luminosity is normalized to the Eddington
luminosity ($L_{\rm Edd}$), the accretion rate will be normalized
to the corresponding Eddington accretion rate ($\dot{M}_{\rm Edd}$).
Lin et al. (2007a, 2009a) reported that XTE~J1701$-$462 revealed clear
photospheric radius expansion during its second and third type-I X-ray bursts.
The empirically determined Eddington luminosity for radius expansion for a
helium burst should be $3.8\times10^{38}$~ergs~s$^{-1}$ (Kuulkers et al. 2003),
which is regarded as the Eddington luminosity of XTE~J1701$-$462 ($L_{\rm Edd}$)
and the corresponding $\dot{M}_{\rm Edd}$ of XTE~J1701$-$462 should
be $3.24\times10^{-8}$~$M_\odot$~yr$^{-1}$.

In a radiation pressure dominated disk, the vertical gravity is balanced by
radiation, as shown in Figure 12. Then the thickness of the inner
disk $z_0 \propto\dot{M}_{\rm disk}$, and thus the density $\rho$ in the inner
disk region is roughly independent of $\dot{M}_{\rm disk}$ (Shakura \& Syunyaev 1973).
Numerical simulations have shown that even for very high accretion
rates, e.g., $\dot{M}_{\rm disk,\ Edd}=3\times 10^2 - 3\times 10^3$, the vertical
thickness of the inflowing part of the inner disk only reaches asymptotically
to about the diameter of the NS, due to the increasing outflow as the total
accretion rate increases (Ohsuga 2007). Therefore, for a constant $\rho$ and
constant vertical thickness of the inflow, $\dot{M}_{\rm disk}\propto v_{\rm r}/v_{\rm ff}$
(note $\dot{M}_{\rm disk}$ throughout this paper describes only the inflow
rate, not the total accretion rate that includes also the unseen outflow rate). We
therefore have
\begin{equation}
 B_{\rm a}=B_0\sqrt{\xi v_{\rm r}/v_{\rm
ff}}\propto\dot{M}_{\rm disk}.
\end{equation}
Substitute equation (6) into equation (1), we get,
\begin{equation}
R_{\rm in} \propto {\dot{M}_{\rm disk}}^{2/7},
\end{equation}
for a radiation pressure dominated disk interacting with a NS's magnetosphere.

The BB component in the spectral models is interpreted as the emission from
the NS surface, so the NS radius can also be derived from the BB temperature
and BB luminosity and the derived NS radius should not be correlated with any
parameter such as the disk accretion rate. Due to the electron scattering on the
NS surface, the observed X-ray spectrum at infinity has a higher temperature than
the effective temperature, i.e., a color correction must be taken. Theoretical
calculations have shown that the color correction factor is between 1.33 to 1.84
(Madej, Joss, \& R{\'o}{\.z}a{\'n}ska 2004). Here we take the value of color
correction factor as $\sim$1.6, since it results in an average value of
around 10~km for the NS radius (a slightly different choice of the color
correction factor will not change any of our conclusions, since we are not
interested in the precise value of the NS radius in this work).

The inferred NS apparent magnetic field strength ($B_{\rm a}$), accretion
rates in the disk and onto the NS ($\dot{M}_{\rm disk}$ and $\dot{M}_{\rm ns}$),
and the NS radius ($R_{\rm ns}$) are listed in Table 5. Figure 13 shows the
correlations between $\dot{M}_{\rm disk}$ and other inferred
parameters, i.e., $R_{\rm in}$, $B_{\rm a}$, $\dot{M}_{\rm ns}$,
and $R_{\rm ns}$. Significance and implications of these quantities and their
evolution will be discussed in the discussion section.

\section{DISCUSSION}

\subsection{The spectral evolution along the HID tracks}

Figure 8 shows the evolution of the fitting parameters and mass accretion
rates along the ``Z'' track and ``$\nu$'' track. Looking at Figure 8, one
can see that there is a transition region (the green region) along each of
the two tracks and the fitting parameters or accretion rates evolve differently
across the transition region. The transition regions of the ``Z'' track
and ``$\nu$'' track are the NB region and the beginning section of the
FB, respectively. This behavior may indicate that the disk structure is
different before and after the transition regions, while the transition
regions connect two different disk states.

As shown by panels C, D, E, and F in Figure 8, along the ``Z'' or ``$\nu$''
track, both the BB flux and BB radius basically remain stable before the
transition region, but they begin to decrease in the transition regions, and
continuously decrease after the transition regions. Both the BB flux and BB
radius are much smaller in FB than in HB, in other words, along the ``Z''
or ``$\nu$'' track the BB component contributes obviously less when the
source ascends the FB. Our results are consistent with those of
Lin, Remillard, \& Homan (2009b). From the BB luminosity, interpreted as the
emission from the NS, we infer the mass accretion rate onto the NS. As shown
by panels G and H in Figure 8, the mass accretion rate onto the NS is
approximately a constant before the transition regions, but it begins to
decrease in the transition regions. Moreover, it is always less than the mass
accretion rate through the inner accretion disk. There are two possibilities
for the obviously small NS mass accretion rate. The first is that the NS is
partially obscured by the accretion disk, so that only part of the NS surface
is visible to the observer; this is very likely unless the accretion disk is
oriented close to face-on. Furthermore, as will be discussed in the next
subsection, the disk structure evolves substantially over these observations
and thus the obscuration to the NS surface also changes accordingly, producing
the observed variations of the inferred NS radius; this is consistent with the
suggestion of Lin, Remillard, \& Homan (2007b, 2009b). This is also consistent
with the frequently observed dipping or flaring during the LFB and FB phase
(Homan et al. 2007a; Shirey, Bradt, \& Levine 1999; Lin, Remillard, \& Homan 2009b).

As shown by panels A and B of Figure 8, along the ``Z'' or ``$\nu$'' track,
both $kT_{\rm in}$ and $kT_{\rm bb}$ do not change remarkably before the transition
regions, and however, they evolve after the transition regions. The disk flux or
total flux increases along each track before the transition regions, and after the
transition regions, it decreases along the ``Z'' track but increases slightly along
the ``$\nu$'' track, as demonstrated in panels C and D of Figure 8. We investigate
the correlations between the luminosities of the thermal components and their
characteristic temperatures, which are shown in the left panels of Figure 10. As
shown in these panels, both the disk and NS do not follow
the $L_{\rm X}\propto T^4$ relationship. It is reasonable, because both the inner
disk radius and BB radius evolve. As shown by the top panel of the right column of
Figure 10, both $R_{\rm in}$ and $R_{\rm bb}$ increase with the PCA intensity in
the HB. The increase of the inner disk radius indicates that during the Z-source
stages the boundary layer expands when the source intensity increases. However,
during the Atoll-source stages, both the disk and NS evolve roughly along
the $L_{\rm X}\propto T^4$ track and both $R_{\rm in}$ and $R_{\rm bb}$ nearly
remain constant (Lin, Remillard, \& Homan 2009b).

It has frequently been assumed that the mass accretion rate ($\dot{M}$) changes
monotonically along the ``Z'' track. For explaining the ``parallel tracks''
phenomenon in NSXBs (M\'endez et al. 1999), van der Klis (2001) proposed that the
frequencies of the kilohertz quasi-periodic oscillations (kHz QPOs) was determined by
the ratio of the disk accretion rate ($\dot{M}_{\rm disk}$) to its own long-term
average ($<\dot{M}_{\rm disk}>$) and the luminosity was determined by $\dot{M}_{\rm disk}$
plus a contribution from $<\dot{M}_{\rm disk}>$. Applying the model of van der Klis (2001)
to XTE~J1701$-$462, Homan et al. (2007a) suggested that the position of the ``Z'' track
is determined by the ratio $\dot{M}_{\rm disk}$/$<\dot{M}_{\rm disk}>$ and the CD shape
evolves as a function of decreasing $\dot{M}$. In this paper, we investigate the evolution
of mass accretion rate along the ``Z'' and ``$\nu$'' tracks. As shown by panels G and H in
Figure 8, along the ``Z'' or ``$\nu$'' track, the disk accretion rate increases before the
transition region, but decreases after the transition region; the NS accretion rate roughly
keeps unvaried before the transition region, and like the disk accretion rate, it decreases
after the transition region. The average disk and NS mass accretion rates of the ``Z'' track
are 7.50~$\dot{M}_{\rm Edd}$ and 0.11~$\dot{M}_{\rm Edd}$, respectively, and those of
the ``$\nu$'' track are 2.09~$\dot{M}_{\rm Edd}$ and 0.01~$\dot{M}_{\rm Edd}$, respectively.
Our results suggest that the mass accretion rates do not monotonically evolve along the ``Z''
track or ``$\nu$'' track in XTE~J1701$-$462 and the source switches from the ``Z'' track to
the ``$\nu$'' track when the mass accretion rates decrease, in agreement with the results
of Homan et al. (2010).

Finally, we point out that none of the fitting parameters monotonically
evolves along the ``Z'' or ``$\nu$'' track, indicating that the HID tracks
are not uniquely determined by only a single fitting parameter.

\subsection{Comparison between our results and others}

In this work, we use the hybrid model of Lin, Remillard, \& Homan (2007b) to
fit the spectra of XTE~J1701$-$462 during a Cyg-like Z episode (the ``Z'' track) and
a Sco-like Z episode (the ``$\nu$'' track). Comparing the unfolded spectra shown in
Figures 3 and 4 with those shown in Figure 13 in the paper of Lin, Remillard, \& Homan
(2009b), one can see that the models in our analyses are consistent with those of theirs.
In order to compare our results with those of Lin, Remillard, \& Homan (2009b), we show
the fraction of CBPL component in Figure 9. Comparing this figure with Figure 12 in their
paper, one can see that our result is consistent with theirs for the Z stages, which is
included in Figure 12 of their paper: during the Z-source stages, the maximum fraction is
approximately 20\%, it decreases with the PCA intensity, and most values of the fraction
are less than 10\%. In Figure 16 in their paper, they showed their spectral fitting results
throughout the Z-source stages. In this figure, they showed the luminosities of thermal 
components (MCD/BB) vs. their characteristic temperatures as well as the PCA intensity, 
also showed the emission sizes of the thermal component vs. the PCA intensity. We show our 
results in Figure 10. Comparing the two figures, we find that our results are in agreement 
with theirs in many aspects, for example, (1) $L_{x}$ vs. $T$ relations deviate from
the $L_{x} \propto T^{4}$ track; (2) in HB, both the MCD luminosity and BB luminosity
increase with the PCA intensity and both $R_{\rm in}$ and $R_{\rm bb}$ also increase with
the PCA intensity; (3) both the BB flux and BB radius decrease dramatically in FB.

Lin, Remillard, \& Homan (2009b) found that in the FB during the Sco-like
Z stages the disk evolved by closely following the $L_{x} \propto T^{4/3}$
track (see section 4.3.2 and Figure 15 of their paper), which they
interpreted as a sign that the inner disk radius is varying under the
condition of a constant $\dot{M}$. We consider the same issue. We fit the
results of the FB of the ``$\nu$'' track by using the least square method.
Contrary to what was found by them, we find that the disk evolves following
the $L_{x} \propto T^{0.45}$ track in the FB during the Sco-like Z stage,
which is demonstrated by the solid line in the bottom panel of the left
column of Figure 10 (see the triangles around the solid line). The difference
between ours and theirs may be due to that in our analyses we estimate the
fluxes in the energy bands lower than the low limit of PCA response matrix
and use the disk luminosity in the whole energy band to
calculate $\dot{M}_{\rm disk}$. The luminosity in the whole energy band is
crucial for inferring $\dot{M}$, especially for deriving $\dot{M}_{\rm disk}$,
because the temperature is below several keV. Moreover, as shown by panel F
in Figure 8, within the FB of the ``$\nu$'' track the inner disk radius is
variable, but the disk accretion rate ($\dot{M}_{\rm disk}$) is not a
constant, as shown by panel H in Figure 8.

Figure 11 shows the correlations between the total luminosity and the
luminosities of the thermal components as well as the disk mass accretion
rate. Baluci\'nska-Church et al. (2010) analyzed the spectra along a
``Z'' track of Cyg X-2 and investigated the same correlations. It is
interesting to compare our results from the ``Z'' track with those of theirs.
As shown in the top panel of Figure 4 in their paper, in the total luminosity
vs. BB luminosity map, the results from the HB form a horizontal line, and in
the corresponding map (panel A in Figure 11), the results from the UHB and HB
also constitute a horizontal line approximately. From NB to FB, the BB
luminosity continuously increases in Cyg X-2, whereas it continuously decreases
from NB, via FB, to LFB in XTE~J1701$-$462. This comparison shows that the
mechanism for producing the BB component is similar in the HBs in the two
sources, but different in the NBs or FBs. As shown in the bottom panel of
Figure 4 in their paper and panel C in Figure 11 here, in the the total
luminosity vs. disk luminosity maps the two sources move along two similar
tracks. However, it is interesting that on the tracks the two sources go along
opposite directions. Assuming the HB (or UHB) to be the beginnings of the tracks,
one can see that Cyg X-2 goes in the clockwise direction, while XTE~J1701$-$462
moves counterclockwise. The similar shapes of the tracks may be due to that the
longitudinal coordinates of the two panels all represent the disk emission. On
the other hand, the contrary evolution directions might be contributed to the
diversity of the mechanism for producing the disk emission. Here, the disk
emission of XTE~J1701$-$462 is interpreted as the model of the multicolor disk
blackbody, while the disk emission of Cyg X-2 was explained by using the
extended accretion disk corona (ADC)
model (Church \& Baluci\'nska-Church 2004; Baluci\'nska-Church et al. 2010)

\subsection{The disk structure evolution}

Figure 13 shows the correlations between $\dot{M}_{\rm disk}$ and several
parameters of the accretion disk and NS, i.e., the inner disk radius ($R_{\rm in}$ in
panel A), the NS apparent magnetic field strength ($B_{\rm a}$ in panel B), the NS
accretion rate (panel C), and NS radius ($R_{\rm ns}$ in panel D), respectively. Panels A
and B show that both $R_{\rm in}$ and $B_{\rm a}$ increase with $\dot{M}_{\rm disk}$. The
dashed lines in panels B and A of Figure 13 are $B_{\rm a}\propto\dot{M}_{\rm disk}$
and $R_{\rm in} \propto {\dot{M}_{\rm disk}}^{2/7}$ given in equations (6) and (7),
respectively; clearly the model predictions agree with data well.

Therefore our results presented here support that the accretion disk is
radiation pressure dominated and $z_0$ increases generally
with $\dot{M}_{\rm disk}$. Considering that the correlations in panels
A and B of Figure 13 hold across a range of $\dot{M}_{\rm disk}$ by a factor
of about ten, we therefore infer that the smallest values of $\xi$ \and $v_{\rm r}/v_{\rm ff}$
are equal to or smaller than 0.1 (because the largest value cannot exceed unity), which
roughly correspond to the thin disk model (Shakura \& Syunyaev 1973). We
therefore suggest that the disk evolves from a thin disk at the
lowest $\dot{M}_{\rm disk}$ to a slim disk at the highest $\dot{M}_{\rm disk}$. Our
conclusion is qualitatively in agreement with the conclusion of Lin, Remillard, \&
Homan (2009b), in which the local Eddington luminosity determines the vertical
structure of the disk.

Panels C and D in Figure 13 show that for $\dot{M}_{\rm disk}$ there are obviously two
branches, i.e., the upper and lower branches. The upper branch corresponds to the
UHB and HB with a constant NS radius and slow increase of $\dot{M}_{\rm ns,\ Edd}$
as $\dot{M}_{\rm disk,\ Edd}$ increases from about 2 to about 10. A rough fitting
gives $\dot{M}_{\rm ns,\ Edd}=0.16 \dot{M}_{\rm disk,\ Edd}^{1/2}$, indicating that
a significant amount of inflowing matter in the inner disk does not reach the NS and
thus must be ejected from the inner disk region as outflow. The lower branch corresponds
to the LFB and FB, and the latter is connected to the HB by the NB. The much
smaller $\dot{M}_{\rm ns,\ Edd}$ and inferred NS radius suggest that very small amount
of inflow matter can reach the NS and we only observe a small fraction of the NS surface.
This suggests a much stronger outflow from the inner disk area and the outflow also blocks
a significant fraction of the NS surface. Inspecting panels A and B in Figure 8, we can
see that $kT_{\rm bb}$ also increases dramatically in FB and LFB, indicating that we
are actually seeing the polar region of the NS, where the accretion column channeled by
the dipole magnetic field causes a hotter radiation area. This picture is consistent with
the frequently observed dipping or flaring phenomenon during the LFB and
FB (Homan et al. 2007a; Shirey, Bradt, \& Levine 1999; Lin, Remillard, \& Homan 2009b).

\subsection{The surface magnetic field strength of the NS in XTE~J1701$-$462}

The NS surface magnetic field strength ($B_0$) is a critical parameter for
understanding the properties of different kinds of NSs. In panel B of Figure 13, the
largest value of $B_{\rm a}$ is about $(1.5-2.3)\times 10^9$ G.
Because $v_{\rm r}/v_{\rm ff}\propto \dot{M}_{\rm disk}$ and $v_{\rm r}/v_{\rm ff}\le 1.0$, we
obtain the lower limit to the surface magnetic field of the NS as $B_0\ge 1.5\times 10^9$ G.
Considering uncertainties in both the data and model, we take $B_0> 1 \times 10^9$ G.

As discussed above, at high $\dot{M}_{\rm disk}$, significant outflow is produced,
reducing significantly $\dot{M}_{\rm ns}$. Shakura \& Syunyaev (1973) suggested that
during the critical and super-Eddington accretion state, the radiation pressure will
make the inner disk region spherized and outflow is thus produced. The spherization
radius is given by,
\begin{equation}
R_{\rm sp} \approx \frac{9}{4}\times10^{6}m\dot{M}_{\rm disk,\ Edd}\mbox{ }{\rm cm},
\end{equation}
where $\dot{M}_{\rm disk,\ Edd}$ is the disk accretion rate, in unit of Eddington
accretion rate and $m$ is the mass of the compact object, in unit of solar mass
$({M_\odot})$. Applying equation (8) to a NSXB and taking 1.4~${M_\odot}$ as the NS
mass, we get,
\begin{equation}
R_{\rm sp} \approx 32\dot{M}_{\rm disk,\ Edd}\mbox{ }{\rm km}.
\end{equation}
Clearly $R_{\rm sp}$ increases faster with $\dot{M}_{\rm disk}$ than $R_{\rm in}$ does ($R_{\rm
in}\propto{\dot{M}_{\rm disk}}^{2/7}$). This implies that at the lowest $\dot{M}_{\rm disk}$, the disk is
terminated by the magnetic pressure of the NS; however at the highest $\dot{M}_{\rm disk}$, the disk would be
terminated (spherized) by the radiation pressure, if $R_{\rm sp}>R_{\rm in}$. Because the highest $\dot{M}_{\rm
disk}$ for the ``Z'' track of HID1 is above the Eddington rate and $R_{\rm in}<30$ km, we have $R_{\rm sp} >
R_{\rm in}$ at the highest $\dot{M}_{\rm disk}$. In this case we have spherical accretion, i.e., $\xi\approx
1.0$. It is easy to show that, $v_{\rm r}/v_{\rm ff}=\sqrt{1-R_{\rm in}/R_{\rm sp}}$. Taking the maximum
$\dot{M}_{\rm disk,\ Edd}\approx 10$, we get $R_{\rm sp}\approx 320$ km. With $R_{\rm in}\approx 28$ km, we get
$v_{\rm r}/v_{\rm ff}\approx 1$. Finally we get the upper limit to $B_0$ as $B_0\le 2.3\times 10^9$~G.
Considering uncertainties in both the data and model, we take $B_0< 3 \times 10^9$ G. We therefore get $(1 <
B_0< 3)\times 10^9$~G, which is much higher than the reported $B_0$$\sim$(0.3--1)$\times10^8$~G of the Atoll
source Aql~X-1 (Zhang, Yu, \& Zhang 1998; Campana 2000), but very close to or just slightly lower than the
reported $B_0= (2.2\pm 0.1)\times10^{9}$~G of the Z source Cyg~X-2 (Focke 1996).

The reported $B_0$ of Aql~X-1 was based on interpreting the observed sudden
spectral transition as due to the ``propeller'' effect, during the flux decay of
an outburst (Zhang, Yu, \& Zhang 1998). In this model, the accretion disk was
assumed to be geometrically thin. However, it is quite possible that the disk at
such a low accretion accretion is an advection dominated accretion flow (ADAF)
and is thus geometrically thick (Zhang, Yu, \& Zhang 1998; Menou et al. 1999).
In this case, less magnetic pressure is needed to cause the assumed ``propeller''
effect for the same accretion rate. Therefore the reported $B_0$ of Aql~X-1 is
only an upper limit, i.e., $B_0\leq$(0.3--1)$\times10^8$~G for the Atoll source
Aql~X-1. The reported $B_0$ of Cyg~X-2 (Focke 1996) was based on interpreting
the observed horizontal-branch oscillations (HBOs) with the beat frequency model
of a spherical accretion (van der Klis et al. 1985); the spherical accretion
assumption tends to under-estimate $B_0$, because more magnetic pressure is
required to balance the gas pressure of a non-spherical accretion of the same
accretion rate, i.e., $B_0\geq(2.2\pm 0.1)\times10^{9}$~G for the Z source
Cyg~X-2. The above values of $B_0$ also agree with the theoretical modeling of
the evolution of NS's surface magnetic field strength due to accretion, resulting
in $\sim$$10^8$~G and $\sim$$10^9$~G for atoll and Z sources, respectively
(Zhang \& Kojima 2006; Zhang 2007). However, Titarchuk, Bradshaw \& Wood (2001)
reported that three NSs, namely, 4U~1728$-$42 (atoll source), GX~340+0 (Z-source),
and Sco~X$-$1 (Z-source), have similar, and very low surface magnetic field
strengths of around $10^{6-7}$~G, based on their magneto-acoustic wave model
for the observed kilohertz Quasi-periodic Oscillations (QPOs) from NSs.

Nevertheless it is possible that the NS surface magnetic field strength in
XTE~J1701$-$462 is between normal atoll and Z sources, although the surface
magnetic field has not been determined reliably for most atoll and Z sources.
Here we suggest that a source in the $\dot{M}$ vs $B_0$ plot may appear as
either a Z source or an atoll source, depending upon the combination of its
NS surface magnetic field strength and $\dot{M}_{\rm disk}$, as shown in
Figure 14; the parameter space above the diagonal line is the Z-source zone
and below that is the atoll-source zone. For a given NSXB, i.e., with a
fixed $B_0$ in the range of approximately between 10$^7$--10$^{10}$ G, it may
change its behaviors from an atoll source to a Z source when
its $\dot{M}_{\rm disk}$ increases until it crosses the diagonal line, or
vice versa, such as the case for XTE~J1701$-$462, which changed its behaviors
from a Z source to an atoll source when its luminosity was sufficiently low,
as just suggested by Lin, Remillard, \& Homan (2009b) and Homan et al. (2010).

\subsection{The hard tail in XTE~J1701$-$462}

Compared with the black hole X-ray binaries (BHXBs), the hard X-ray emission 
is not frequently observed in NSXBs due to the emission from the surface of 
the NS, absence of a high temperature corona, and seldom jets. The persistent 
Z sources present complicated behaviors of hard tail and its origin is still 
debated. A possible explanation was that the hard tail in NSXBs was owing to 
the Compton up-scattering of soft X-ray photons by the high energy nonthermal 
electrons from the jets (Di Salvo et al. 2000; Iaria et al. 2001). In addition 
to this explanation, some other interpretations have been proposed, such as 
due to synchrotron emission (D'Amico et al. 2001; Iaria et al. 2002). Since 
the disk structure of a NSXB is similar to that of a BHXB, it is plausible to 
use the mechanisms for the power-law component of BHXBs to explain the hard 
tail of NSXBs, e.g., due to repeated Compton scattering of seed photons from 
the disk by the thermal/nonthermal electrons in a corona (Gierli\'nski et al. 
1999; Poutanen \& Coppi 1998), strong disk Comptonization (Kubota, Makishima, 
\& Ebisawa 2001a), and magnetic turbulence (Kontar, Hannah, \& Bian 2011), or 
resulted from an optically-thick ADAF (Kubota, Makishima, \& Ebisawa 2001b).

Farinelli et al. (2007, 2008, 2009) proposed that the hard X-ray emission in NSXBs
originated from the Comptonization of soft photons by matter undergoing relativistic
bulk-motion in the region near the NS. In this work, in order to investigate the origin 
of the hard tail in XTE~J1701$-$462, we use the physical model of Farinelli et al. (2007) 
to fit the UHB and HB spectra of XTE~J1701$-$462, which exhibit hard tails, as shown in 
Figures 3 and 4. This model consists of two components: one is the CompTT model, 
developed by Titarchuk (1994), which describes the Comptonization of soft photons in 
a hot plasma near the inner disk; the other is the BMC model (Titarchuk, Mastichiadis, 
\& Kylafis 1996, 1997), which describes the Comptonization of soft photons by matter 
undergoing relativistic bulk-motion near the NS. The $\chi^2~(dof)$ listed in Table 4 
and the residual distribution displayed in Figure 7 show that the spectra can be fit 
statistically well. The fitting results are listed in Table 4, which are consistent with 
those of Farinelli et al. (2007). Due to absence of the spectra lower than $\sim$3~keV, 
the seed photon temperature of CompTT cannot be constrained well, so it is fixed to 0.5 keV. 
The seed photon temperature of BMC is over 1 keV, which is higher than the seed photon 
temperature of CompTT, indicating that the seed photons for CompTT could come from cooler 
emission regions such as the disk or the transition layer near the inner disk, while the 
seed photons for BMC should come from hotter emission region such as the region near the 
NS. In our analyses, we find that the two parameters $\alpha$ and $\log(A)$ of BMC cannot 
be determined at the same time, so we fix $\alpha$. The values of $\log(A)$ are negative 
and the Comptonization factor $f$ spans a range of $\sim$(0.5--33)\%. The electron 
temperature of CompTT is over $\sim$2.5~keV and optical depth is over 10. Moreover, as 
shown by the unfolded spectra in Figure 5, the BMC extends to higher energy bands than 
the CompTT and the convolution Comptonization component in BMC accounts for the hard tail. 
Our analyses suggest that the spectra of XTE~J1701$-$462 which have the hard tail can be 
fit by the CompTT+BMC+LINE model statistically well and the Comptonization of soft photons 
by matter undergoing relativistic bulk-motion near the NS could be an alternative mechanism 
for producing the hard tail in XTE~J1701$-$462

As shown by Figures 3 and 4, the spectra in the UHB and HB are fit by the
MCD+BB+CBPL+LINE model and the spectra in the higher energy bands are mainly
fit by the CBPL component. The CBPL component in this model was interpreted
as a component from a very weakly Comptonized disk (Lin, Remillard, \& Homan 2007b, 2009b).
The $\chi^2~(dof)$ listed in Table 3 and the residual distribution demonstrated in Figure 6 
show that the fitting is statistically good, so the weak
Comptonization taking place in a very weakly Comptonized disk might be another possible process
responsible for the hard tail in this source.

Di Salvo et al. (2000) used the BB+CompTT+PL+LINE model to fit the spectra of GX 17+2 in
0.1--200 keV and detected the hard tail in the HB. They used the power law (PL) component 
in this model to account for the hard tail and suggested an alternative possibility that 
this observed PL component originated from Compton up-scattering of the soft X-ray photons 
by the high-energy nonthermal electrons from a jet. In our anlyses, we try to fit the UHB 
and HB spectra of XTE~J1701$-$462 by using this model and find that the fitting is 
statistically acceptable too. Therefore, it cannot be excluded that the hard tail 
of XTE~J1701$-$462 is also associated with a jet.

\section{CONCLUSION}

We have investigated the spectral evolution along a ``Z'' track and
a ``$\nu$'' track of the first transient Z source XTE~J1701$-$462, inferred
its mass accretion rates and NS surface magnetic field strength, and studied its
hard tail. We summarize our results as follows:

1. The low-energy spectra can be fit by a MCD+BB+LINE model, while the broadband
spectra can be fit by a MCD+BB+CBPL+LINE model statistically well and the CBPL component, 
accounting for the emission from a very weakly Comptonized accretion disk, contributes 
significantly. Along the``Z'' or ``$\nu$'' track, the BB emission roughly keeps unvaried 
until the HB/NB vertex, but it begins to decrease in the NB, and decreases dramatically 
in the FB.

2. The observed positive correlation between the inner accretion disk radius and
disk accretion rate ($R_{\rm in}\propto \dot{M}_{\rm disk}^{2/7}$) is consistent with
the prediction of a simple model of the interaction of a radiation pressure dominated
accretion disk with the magnetosphere of a NS.

3. The radiation pressure thickens the disk considerably, and also produces
significant outflows.

4. The significantly increased outflow from the inner disk regions in LFB
and FB blocks much of the solid angle to the NS surface and we can only
see the hotter polar region of the NS.

5. The surface dipole magnetic field strength of the NS, inferred from the interaction
between the magnetosphere and the radiation pressure dominated accretion disk,
is $\sim$(1--3)$\times10^9$~G, possibly between that of normal Z and atoll sources.

6. A significant hard tail is detected in HB, which can be interpreted by several
alternative mechanisms including the bulk-motion Comptonization near the NS, the weak
Comptonization taking place in a very weakly Comptonized disk, or inverse Compton
scattering of soft X-ray photons by the higher-energy electrons from a jet, etc..

\acknowledgments

We thank the anonymous referees for their constructive criticism and suggestions,
that have allowed us to improve significantly the presentation of this paper. We
appreciate discussions with Roberto Soria, Feng Yuan and Shanshan Weng. This
research has made use of data obtained through the High Energy Astrophysics Science
Archive Research Center (HEASARC) On-line Service, provided by NASA/Goddard Space
Flight Center (GSFC). This work is partially supported by the Natural Science
Foundation of Xinjiang Uygur Autonomous Region of China (Program 200821164) and
the program of the Light in Chinese Western Region (LCWR) under grant LHXZ 200802
provided by Chinese Academy of Sciences (CAS), and supported by National Basic
Research Program of China (973 Program 2009CB824800), National Science Foundation
of China under grant No. 10821061, 10733010, and 10725313.

\clearpage

\begin{table}
\caption{Log of used RXTE observations for XTE~J1701$-$462 in HID1}
\begin{tabular}{lccccc}
\tableline\tableline
                &             &               &      PCA    &    HEXTE    &    \\
                &             &               &   Good Time &  Good Time  &    \\
                &    Date     & $\rm MJD^{a}$ &   Interval  &   Interval  &    \\
OBSID           &    (UTC)    &    (Day)      &     (ks)    &   (ks)      & HID position \\
\hline
91106-01-04-00  & 2006 Jan 21 &   53756.6116  &     4.416   &    3.780    & FB, NB \\
91106-01-05-00  & 2006 Jan 21 &   53756.7078  &     2.096   &    2.018    & FB, NB \\
91106-01-06-00  & 2006 Jan 21 &   53756.7777  &     5.008   &    2.626    & FB, NB \\
91106-01-07-00  & 2006 Jan 22 &   53757.7581  &     7.088   &    3.750    & NB, HB \\
91106-01-08-00  & 2006 Jan 23 &   53758.1481  &     1.216   &    0.840    & NB \\
91106-01-10-00  & 2006 Jan 24 &   53759.5081  &     3.520   &    2.282    & NB \\
91106-01-11-00  & 2006 Jan 24 &   53759.6240  &     4.560   &    3.430    & NB \\
91106-01-12-00  & 2006 Jan 24 &   53759.7659  &     7.360   &    6.325     & NB, HB \\
91106-01-13-00  & 2006 Jan 25 &   53760.0295  &     3.584   &    3.158    & FB, NB \\
91106-02-01-00  & 2006 Jan 25 &   53760.6832  &     3.920   &    0.000    & FB, NB \\
91106-02-01-01  & 2006 Jan 25 &   53760.7492  &     3.056   &    0.000    & FB, NB \\
91106-02-01-02  & 2006 Jan 25 &   53760.9440  &     3.522   &    1.600    & LFB, FB \\
91106-02-01-03  & 2006 Jan 26 &   53761.5890  &     4.256   &    3.528    & HB \\
91106-02-02-00  & 2006 Jan 27 &   53762.2641  &     2.864   &    2.020    & HB \\
91106-02-02-01  & 2006 Jan 27 &   53762.5870  &     13.84   &    9.550    & HB \\
91106-02-02-02  & 2006 Jan 28 &   53763.5532  &     2.448   &    2.080    & UHB \\
91106-02-02-03  & 2006 Jan 28 &   53763.8380  &     2.096   &    2.010    & UHB \\
91106-02-02-04  & 2006 Jan 29 &   53764.0921  &     3.312   &    2.835    & UHB \\
91106-02-02-05  & 2006 Jan 29 &   53764.3551  &     3.488   &    2.740    & UHB \\
91106-02-02-06  & 2006 Jan 29 &   53764.4064  &     3.920   &    3.600    & UHB \\
91106-02-02-07  & 2006 Jan 29 &   53764.5342  &     2.336   &    1.800    & UHB \\
91106-02-02-08  & 2006 Jan 30 &   53765.5477  &     2.336   &    1.750    & UHB \\
91106-02-02-09  & 2006 Jan 30 &   53765.9608  &     3.792   &    3.320    & UHB, HB \\
91106-02-02-10  & 2006 Jan 31 &   53766.8987  &     6.448   &    6.240    & HB \\
\tableline
\end{tabular}
\tablecomments{Goog time refers to the observation time with offset (pointing position) less than 0.02 arcsecond
and elv (elevation angle) larger than 10 degrees.}
\tablenotetext{a}{Start of observation, MJD=JD--2,400,000.5}
\end{table}

\begin{table}
\caption{Log of used RXTE observations for XTE~J1701$-$462 in HID2}
\begin{tabular}{lccccc}
\tableline\tableline
                &             &               &     PCA     &    HEXTE    &    \\
                &             &               &  Good Time  &  Good Time  &    \\
                &    Date     & $\rm MJD^{a}$ &   Interval  &   Interval  &    \\

OBSID           &    (UTC)    &    (Day)      &     (ks)    &   (ks)      & HID position \\
\hline
91442-01-03-09  & 2006 Mar 2  &   53796.6620  &     1.696   &    1.680    & FB \\
91442-01-03-10  & 2006 Mar 2  &   53796.7131  &     2.928   &    2.900    & FB \\
92405-01-01-00  & 2006 Mar 3  &   53797.7550  &     6.946   &    6.290    & FB \\
92405-01-01-01  & 2006 Mar 4  &   53798.6828  &     5.616   &    5.460    & NB \\
92405-01-01-02  & 2006 Mar 5  &   53799.7864  &     3.264   &    3.160    & HB \\
92405-01-01-03  & 2006 Mar 6  &   53800.7030  &     3.280   &    3.220    & FB \\
92405-01-01-04  & 2006 Mar 7  &   53801.6850  &     3.280   &    3.210    & HB \\
\tableline
\end{tabular}
\tablecomments{Goog time refers to the observation time with offset (pointing position) less than 0.02 arcsecond
and elv (elevation angle) larger than 10 degrees.}
\tablenotetext{a}{Start of observation, MJD=JD--2,400,000.5}
\end{table}

\clearpage

\begin{table}
\footnotesize \caption{PCA+HEXTE spectral fitting parameters for HID regions in XTE~J1701$-$462, using the
MCD+BB+CBPL+LINE model for the UHB and HB spectra, and the MCD+BB+LINE model for the NB, FB, and LFB spectra.}
\begin{tabular}{lccccccccc}
\tableline\tableline
HID region & $kT_{\rm in}$ (keV) & $^{a}R_{\rm in}\sqrt{\cos\theta}$ & $^{b}flux_{\rm diskbb}$ & $E_{\rm Fe}$ (keV) & $^{c}flux_{\rm Fe}$ & $kT_{\rm bb}$ (keV) & $^{d}R_{\rm bb}$ (km) &$^{e}flux_{\rm bb}$ & $\chi^2~(dof)$ \\
\tableline 1 (HID1) &
    $1.60_{-0.08}^{+0.08}$  & $7.48_{-0.50}^{+0.61}$ &
    $11.17_{-1.34}^{+1.67}$ & $6.61_{-0.13}^{+0.14}$ &
    $1.23_{-0.23}^{+0.26}$  & $2.42_{-0.09}^{+0.09}$ &
    $3.19_{-0.28}^{+0.35}$  & $5.04_{-1.10}^{+0.89}$ &
             33.76(42)      \\
\tableline 2 &
    $1.46_{-0.06}^{+0.06}$  & $11.74_{-0.70}^{+0.90}$ &
    $13.72_{-1.64}^{+2.11}$  & $6.58_{-0.15}^{+0.15}$ &
    $1.04_{-0.27}^{+0.28}$  & $2.33_{-0.09}^{+0.09}$  &
    $3.42_{-0.30}^{+0.36}$  & $5.05_{-1.07}^{+0.88}$  &
             26.94(41)     \\
\tableline 3 &
    $1.43_{-0.05}^{+0.05}$  & $11.81_{-0.65}^{+0.82}$ &
    $16.49_{-1.81}^{+2.30}$ & $6.50_{-0.15}^{+0.16}$  &
    $1.14_{-0.30}^{+0.31}$  & $2.31_{-0.01}^{+0.01}$  &
    $3.61_{-0.29}^{+0.36}$  & $5.32_{-0.86}^{+1.07}$  &
             30.73(41)     \\
\tableline 4 &
    $1.46_{-0.06}^{+0.05}$  & $12.57_{-0.69}^{+0.89}$ &
    $20.12_{-2.22}^{+2.83}$ & $6.52_{-0.19}^{+0.20}$  &
    $1.10_{-0.36}^{+0.37}$  & $2.30_{-0.09}^{+0.09}$  &
    $3.81_{-0.35}^{+0.47}$  & $5.89_{-1.10}^{+1.44}$  &
             39.36(41)     \\
\tableline 5 &
    $1.47_{-0.05}^{+0.05}$  & $13.06_{-0.67}^{+0.85}$ &
    $22.68_{-2.32}^{+2.96}$ & $6.50_{-0.21}^{+0.22}$  &
    $1.07_{-0.39}^{+0.40}$  & $2.31_{-0.09}^{+0.08}$  &
    $3.96_{-0.36}^{+0.46}$  & $6.49_{-1.17}^{+1.50}$  &
             36.81(41)     \\
\tableline 6 &
    $1.50_{-0.06}^{+0.05}$   & $13.27_{-0.68}^{+0.87}$ &
    $25.01_{-2.55}^{+3.27}$  & $6.51_{-0.22}^{+0.23}$  &
    $1.11_{-0.42}^{+0.43}$   & $2.30_{-0.08}^{+0.09}$  &
    $4.12_{-0.39}^{+0.50}$   & $6.90_{-1.30}^{+1.67}$  &
             42.22(40)      \\
\tableline 7 &
    $1.52_{-0.05}^{+0.05}$  & $13.47_{-0.68}^{+0.86}$ &
    $27.07_{-2.72}^{+3.45}$ & $6.49_{-0.25}^{+0.26}$  &
    $1.15_{-0.46}^{+0.47}$  & $2.29_{-0.08}^{+0.08}$  &
    $4.27_{-0.40}^{+0.52}$  & $7.31_{-1.36}^{+1.79}$  &
             28.35(41)     \\
\tableline 8 &
    $1.55_{-0.06}^{+0.05}$  & $13.45_{-0.68}^{+0.89}$ &
    $29.57_{-3.00}^{+3.78}$ & $6.48_{-0.26}^{+0.28}$  &
    $1.20_{-0.50}^{+0.52}$  & $2.32_{-0.09}^{+0.09}$  &
    $4.24_{-0.42}^{+0.59}$  & $7.55_{-1.51}^{+2.11}$  &
             26.36(40)     \\
\tableline 9 &
    $1.56_{-0.06}^{+0.04}$  & $14.00_{-0.61}^{+0.88}$ &
    $33.47_{-2.92}^{+4.22}$ & $6.45_{-0.27}^{+0.28}$  &
    $1.20_{-0.52}^{+0.56}$  & $2.33_{-0.12}^{+0.07}$  &
    $4.15_{-0.38}^{+0.74}$  & $7.38_{-1.33}^{+2.62}$  &
             30.35(40)     \\
\tableline 10 &
    $1.55_{-0.05}^{+0.04}$  & $14.59_{-0.65}^{+0.87}$ &
    $34.70_{-3.10}^{+4.14}$ & $6.51_{-0.25}^{+0.26}$  &
    $1.31_{-0.49}^{+0.53}$  & $2.26_{-0.10}^{+0.09}$  &
    $4.14_{-0.47}^{+0.71}$  & $6.50_{-1.46}^{+2.24}$  &
             37.64(41)     \\
\tableline 11 &
    $1.52_{-0.03}^{+0.03}$  & $15.19_{-0.57}^{+0.65}$ &
    $34.81_{-2.63}^{+2.99}$ & $6.45_{-0.27}^{+0.27}$  &
    $1.06_{-0.46}^{+0.46}$  & $2.27_{-0.05}^{+0.06}$  &
    $3.66_{-0.32}^{+0.36}$  & $5.15_{-0.90}^{+1.01}$  &
             31.24(39)     \\
\tableline 12 &
    $1.52_{-0.03}^{+0.03}$  & $15.19_{-0.49}^{+0.55}$ &
    $34.28_{-2.20}^{+2.47}$ & $6.56_{-0.28}^{+0.29}$  &
    $0.82_{-0.39}^{+0.38}$  & $2.32_{-0.06}^{+0.06}$  &
    $2.97_{-0.27}^{+0.31}$  & $3.71_{-0.68}^{+0.78}$  &
             36.28(39)     \\
\tableline 13 &
    $1.44_{-0.05}^{+0.04}$  & $16.37_{-0.71}^{+1.06}$ &
    $32.82_{-2.84}^{+4.27}$ & $6.47_{-0.23}^{+0.23}$  &
    $1.00_{-0.40}^{+0.42}$  & $2.11_{-0.14}^{+0.13}$  &
    $3.43_{-0.57}^{+1.01}$  & $3.37_{-1.11}^{+2.12}$  &
             26.88(39)     \\
\tableline 14 &
    $1.47_{-0.02}^{+0.02}$  & $15.68_{-0.38}^{+0.43}$ &
    $32.38_{-1.57}^{+1.77}$ & $6.51_{-0.28}^{+0.29}$  &
    $0.59_{-0.31}^{+0.33}$  & $2.33_{-0.07}^{+0.08}$  &
    $2.01_{-0.22}^{+0.28}$  & $1.74_{-0.39}^{+0.48}$  &
            32.28(40)      \\
\tableline 15 &
    $1.44_{-0.02}^{+0.02}$  & $16.01_{-0.38}^{+0.43}$ &
    $30.96_{-1.48}^{+1.70}$ & $6.56_{-0.25}^{+0.27}$  &
    $0.59_{-0.28}^{+0.29}$  & $2.36_{-0.11}^{+0.12}$  &
    $1.59_{-0.24}^{+0.33}$  & $1.13_{-0.34}^{+0.47}$  &
            42.60(37)     \\
\tableline 16 &
    $1.42_{-0.02}^{+0.02}$  & $15.61_{-0.52}^{+0.62}$ &
    $28.19_{-1.90}^{+2.23}$ & $6.60_{-0.36}^{+0.45}$  &
    $0.97_{-0.51}^{+0.55}$  & $2.53_{-0.18}^{+0.22}$  &
    $1.05_{-0.22}^{+0.37}$  & $0.66_{-0.27}^{+0.46}$  &
             36.65(37)     \\
\tableline 17 &
    $1.42_{-0.02}^{+0.02}$  & $15.06_{-0.65}^{+0.51}$ &
    $25.84_{-2.25}^{+1.72}$ & $7.83_{-1.35}^{+0.56}$  &
    $0.76_{-0.35}^{+0.45}$  & $2.83_{-0.06}^{+0.05}$  &
    $0.62_{-0.11}^{+0.15}$  & $0.36_{-0.10}^{+0.15}$  &
             37.67(33)     \\
\tableline 18 &
    $1.34_{-0.02}^{+0.02}$  & $14.04_{-0.49}^{+0.66}$ &
    $18.58_{-0.77}^{+1.08}$ & $4.94_{-0.07}^{+0.07}$  &
    $0.39_{-0.08}^{+0.10}$  & $3.24_{-0.01}^{+0.01}$  &
    $0.80_{-0.38}^{+0.30}$  & $0.20_{-0.08}^{+0.09}$  &
             15.92(22)     \\
\tableline\tableline 1(HID2) &
    $1.48_{-0.07}^{+0.07}$  & $8.40_{-0.58}^{+0.72}$  &
    $9.44_{-1.31}^{+1.61}$  & $6.54_{-0.20}^{+0.21}$  &
    $0.66_{-0.23}^{+0.24}$  & $2.43_{-0.09}^{+0.09}$  &
    $3.06_{-0.24}^{+0.28}$  & $4.76_{-0.74}^{+0.88}$  &
             32.93(39)      \\
\tableline 2 &
    $1.46_{-0.07}^{+0.07}$  & $8.97_{-0.60}^{+0.78}$  &
    $10.37_{-1.39}^{+1.81}$ & $6.56_{-0.17}^{+0.18}$  &
    $0.86_{-0.25}^{+0.25}$  & $2.36_{-0.09}^{+0.10}$  &
    $3.26_{-0.28}^{+0.33}$  & $4.75_{-0.82}^{+0.96}$  &
             27.73(39)      \\
\tableline 3 &
    $1.58_{-0.05}^{+0.05}$  & $8.77_{-0.45}^{+0.51}$  &
    $13.61_{-1.39}^{+1.59}$ & $6.58_{-0.23}^{+0.24}$  &
    $0.61_{-0.23}^{+0.23}$  & $2.51_{-0.05}^{+0.05}$  &
    $2.67_{-0.19}^{+0.20}$  & $4.08_{-0.58}^{+0.61}$  &
             34.25(39)      \\
\tableline 4 &
    $1.58_{-0.04}^{+0.04}$   & $8.89_{-0.40}^{+0.45}$  &
    $13.90_{-1.25}^{+1.41}$  & $6.63_{-0.18}^{+0.19}$  &
    $0.76_{-0.21}^{+0.22}$   & $2.51_{-0.05}^{+0.06}$  &
    $2.44_{-0.18}^{+0.19}$   & $3.42_{-0.49}^{+0.53}$  &
             42.30(40)      \\
\tableline 5 &
    $1.63_{-0.04}^{+0.04}$  & $8.75_{-0.35}^{+0.40}$  &
    $15.06_{-1.20}^{+1.37}$ & $6.57_{-0.14}^{+0.14}$  &
    $1.02_{-0.21}^{+0.22}$  & $2.47_{-0.07}^{+0.08}$  &
    $2.09_{-0.21}^{+0.24}$  & $2.37_{-0.48}^{+0.55}$  &
             35.44(39)      \\
\tableline 6 &
    $1.51_{-0.04}^{+0.04}$  & $10.50_{-0.55}^{+0.66}$ &
    $16.16_{-1.70}^{+2.02}$ & $6.63_{-0.25}^{+0.26}$  &
    $1.03_{-0.34}^{+0.36}$  & $2.47_{-0.09}^{+0.10}$  &
    $1.98_{-0.24}^{+0.29}$  & $2.09_{-0.50}^{+0.61}$  &
             37.17(38)      \\
\tableline 7 &
    $1.87_{-0.05}^{+0.04}$  & $6.79_{-0.25}^{+0.30}$ &
    $15.95_{-1.17}^{+1.42}$ & $6.56_{-0.15}^{+0.15}$ &
    $1.07_{-0.23}^{+0.23}$  & $2.79_{-0.14}^{+0.17}$ &
    $1.38_{-0.23}^{+0.32}$  & $1.67_{-0.56}^{+0.77}$ &
             24.35(39)      \\
\tableline 8 &
    $2.11_{-0.04}^{+0.03}$  & $5.50_{-0.14}^{+0.17}$ &
    $16.93_{-0.87}^{+1.05}$ & $6.50_{-0.17}^{+0.18}$ &
    $0.95_{-0.24}^{+0.24}$  & $3.44_{-0.26}^{+0.32}$ &
    $0.72_{-0.15}^{+0.24}$  & $1.06_{-0.43}^{+0.70}$ &
             30.36(38)     \\
\tableline 9 &
    $2.28_{-0.04}^{+0.03}$  & $4.76_{-0.12}^{+0.16}$ &
    $17.16_{-0.86}^{+1.14}$ & $6.61_{-0.19}^{+0.20}$ &
    $0.93_{-0.24}^{+0.25}$  & $3.64_{-0.36}^{+0.44}$ &
    $0.62_{-0.15}^{+0.29}$  & $0.98_{-0.46}^{+0.91}$ &
             36.87(39)     \\
\tableline 10 &
    $2.52_{-0.11}^{+0.06}$  & $4.06_{-0.16}^{+0.30}$  &
    $18.70_{-1.44}^{+2.74}$ & $6.64_{-0.19}^{+0.21}$  &
    $1.09_{-0.28}^{+0.29}$  & $3.56_{-0.50}^{+0.71}$  &
    $0.74_{-0.25}^{+0.78}$  & $1.27_{-0.85}^{+0.70}$  &
             43.75(40)     \\
\tableline 11 &
    $2.87_{-0.12}^{+0.05}$  & $3.34_{-0.09}^{+0.23}$  &
    $21.32_{-1.13}^{+2.88}$ & $6.72_{-0.25}^{+0.30}$  &
    $1.00_{-0.31}^{+0.16}$  & $4.50_{-1.16}^{+2.3}$   &
    $0.35_{-0.15}^{+0.16}$  & $0.31_{-0.61}^{+0.53}$  &
             32.65(39)     \\
\tableline
\end{tabular}
\tablecomments{Errors quoted are 90\% confidence limits for the fitting parameters ($\Delta\chi^2=2.7$). The
interstellar hydrogen column and gaussian width of the line are fixed to $2.0\times10^{22}$~cm$^{-2}$ and 0.3
keV, respectively (Lin, Remillard, \& Homan 2009b).}
\tablenotetext{a}{Inner radius of the accretion disk (times $\sqrt{\cos\theta}$, where $\theta$ is the angle between the normal to the disk and the line of the sight) for a distance of 8.8~kpc to the source, in unit of km}
\tablenotetext{b}{The unabsorbed diskbb (MCD) flux, in 0.01--100 keV, in units of $10^{-9}$~ergs~cm$^{-2}$~s$^{-1}$}
\tablenotetext{c}{The unabsorbed flux of the iron emission line, in units of $10^{-2}$~photons~cm$^{-2}$~s$^{-1}$}
\tablenotetext{d}{BB radius for a distance of 8.8~kpc to the source}
\tablenotetext{e}{The unabsorbed BB flux in 0.01--100 keV, in units of $10^{-9}$~ergs~cm$^{-2}$~s$^{-1}$}
\end{table}

\clearpage

\begin{table}
\scriptsize
\caption{PCA+HEXTE spectral fitting parameters for the HID regions of the UHB and HBs in XTE~J1701$-$462, using a model consisting of the BMC+CompTT model plus a LINE.}
\begin{tabular}{lccccccccccccc}
\tableline\tableline
 & \multicolumn{4}{c}{BMC} & & \multicolumn{3}{c}{CompTT} & & \multicolumn{2}{c}{Line}\\
\cline{2-5} \cline{7-9} \cline{11-12}
 & $kT_{\rm bb}$ & $^{a}\alpha$ & $^{b}log{\rm (A)}$ & $N_{\rm bmc}$ & & $kT_{\rm e}$ & $^{c}\tau$ & $N_{\rm comptt}$ & & $E_{\rm Fe}$ & $N_{\rm Fe}$ &  &  \\
HID region & (keV) &  & & $(\times10^{-2})$ & & (keV) & & & & (keV) & $(\times10^{-2})$ & $^{d}f$ & $\chi^2~(dof)$\\
\tableline 1 (HID1) &
    $1.43_{-0.15}^{+0.19}$  &         0.01(fixed)     &
    $-0.44_{-0.24}^{+0.19}$ & $2.72_{-0.46}^{+0.46}$  &&
    $2.92_{-0.09}^{+0.13}$  & $12.17_{-0.83}^{+0.70}$ &
    $1.15_{-0.06}^{+0.06}$  &&$6.53_{-0.13}^{+0.14}$  &
    $1.33_{-0.27}^{+0.28}$  &           26.64\%       &
            32.86(42)       \\
\tableline 2 &
    $1.19_{-0.09}^{+0.13}$  &         0.01(fixed)     &
    $-0.36_{-0.22}^{+0.21}$ & $3.40_{-0.67}^{+0.81}$  &&
    $2.77_{-0.07}^{+0.08}$  & $11.70_{-0.60}^{+0.67}$ &
    $1.48_{-0.11}^{+0.10}$  &&$6.50_{-0.15}^{+0.15}$  &
    $1.13_{-0.28}^{+0.29}$  &           30.39\%       &
            22.62(41)       \\
\tableline 3 &
    $1.09_{-0.07}^{+0.09}$  &         0.10(fixed)     &
    $-1.36_{-0.22}^{+0.21}$ & $3.53_{-0.93}^{+1.12}$  &&
    $2.65_{-0.07}^{+0.08}$  & $12.02_{-0.67}^{+0.83}$ &
    $1.68_{-0.16}^{+0.14}$  &&$6.50_{-0.15}^{+0.16}$  &
    $1.21_{-0.31}^{+0.32}$  &           4.18\%        &
            30.68(41)       \\
\tableline 4 &
    $1.11_{-0.07}^{+0.08}$  &         0.01(fixed)     &
    $-0.91_{-0.26}^{+0.33}$ & $5.53_{-1.24}^{+1.46}$  &&
    $2.63_{-0.08}^{+0.09}$  & $12.12_{-0.78}^{+0.97}$ &
    $1.90_{-0.20}^{+0.17}$  &&$6.48_{-0.20}^{+0.21}$  &
    $1.12_{-0.38}^{+0.39}$  &           10.95\%       &
            40.64(41)       \\
\tableline 5 &
    $1.10_{-0.06}^{+0.07}$  &         1.64(fixed)     &
    $-1.30_{-0.26}^{+0.19}$ & $5.86_{-1.22}^{+1.42}$  &&
    $2.59_{-0.08}^{+0.09}$  & $12.27_{-0.70}^{+0.86}$ &
    $2.13_{-0.20}^{+0.17}$  &&$6.48_{-0.20}^{+0.21}$  &
    $1.12_{-0.22}^{+0.20}$  &            4.77\%       &
            34.61(41)       \\
\tableline 6 &
    $1.08_{-0.07}^{+0.08}$  &         2.95(fixed)     &
    $-0.30_{-0.16}^{+0.15}$ & $7.80_{-1.51}^{+1.91}$  &&
    $2.53_{-0.08}^{+0.08}$  & $12.47_{-0.80}^{+1.09}$ &
    $2.23_{-0.29}^{+0.24}$  &&$6.48_{-0.22}^{+0.25}$  &
    $1.15_{-0.45}^{+0.46}$  &            33.39\%      &
            40.41(40)       \\
\tableline 7 &
    $1.09_{-0.06}^{+0.08}$  &         0.30(fixed)     &
    $-0.61_{-0.21}^{+0.18}$ & $8.39_{-1.57}^{+1.91}$  &&
    $2.45_{-0.08}^{+0.08}$  & $13.06_{-0.89}^{+1.19}$ &
    $2.46_{-0.31}^{+0.27}$  &&$6.48_{-0.26}^{+0.28}$  &
    $1.15_{-0.49}^{+0.51}$  &            19.71\%      &
            27.68(41)       \\
\tableline 8 &
    $1.11_{-0.07}^{+0.08}$  &         1.93(fixed)     &
    $-1.00_{-0.29}^{+0.21}$ & $9.13_{-1.61}^{+1.96}$  &&
    $2.44_{-0.09}^{+0.10}$  & $13.30_{-0.99}^{+1.30}$ &
    $2.61_{-0.09}^{+0.10}$  &&$6.48_{-0.28}^{+0.31}$  &
    $1.16_{-0.53}^{+0.56}$  &            9.09\%       &
            26.00(40)       \\
\tableline 9 &
    $1.16_{-0.05}^{+0.06}$  &         0.33(fixed)     &
    $-2.36_{-0.47}^{+0.24}$ & $9.88_{-1.43}^{+1.66}$  &&
    $2.49_{-0.06}^{+0.07}$  & $12.64_{-0.74}^{+0.88}$ &
    $2.89_{-0.29}^{+0.26}$  &&$6.41_{-0.28}^{+0.30}$  &
    $1.16_{-0.53}^{+0.56}$  &            0.43\%       &
            29.04(40)       \\
\tableline\tableline 1(HID2) &
    $1.14_{-0.13}^{+0.24}$  &        0.30(fixed)      &
    $-1.14_{-0.41}^{+0.36}$ & $1.49_{-0.61}^{+0.86}$  &&
    $2.92_{-0.10}^{+0.12}$  & $12.15_{-0.76}^{+0.95}$ &
    $1.10_{-0.24}^{+0.28}$  &&$6.48_{-0.20}^{+0.21}$  &
    $0.69_{-0.24}^{+0.25}$  &            6.76\%       &
               30.15(39)    \\
\tableline 2 &
    $1.20_{-0.11}^{+0.16}$  &        0.01(fixed)      &
    $-0.60_{-0.34}^{+0.33}$ & $2.16_{-0.75}^{+0.86}$  &&
    $2.88_{-0.07}^{+0.09}$  & $11.92_{-0.60}^{+0.71}$ &
    $1.19_{-0.09}^{+0.08}$  &&$6.50_{-0.18}^{+0.19}$  &
    $0.88_{-0.25}^{+0.26}$  &            20.08\%      &
               25.99(39)    \\
\tableline
\end{tabular}
\tablecomments{Errors quoted are 90\% confidence limits for the fitting parameters ($\Delta\chi^2=2.7$). The
interstellar hydrogen column and gaussian width of the line are fixed to $2.0\times10^{22}$~cm$^{-2}$ and 0.3
keV, respectively (Lin, Remillard, \& Homan 2009b). The seed photon temperature of CompTT is fixed to 0.5 keV.}
\tablenotetext{a}{Energy spectral index of BMC ($\Gamma=\alpha+1$)}
\tablenotetext{b}{The logarithm of the "A" parameter, indicating the covering of BB by Compt cloud for MBC}
\tablenotetext{c}{The optical depth of the scattering cloud for CompTT, assuming a spherical geometry of the Compt cloud}
\tablenotetext{d}{Comptnization fraction: $f=A/(A+1)$}
\end{table}

\clearpage

\begin{table}
\footnotesize \caption{The estimated NS apparent magnetic field strength and $\dot{M}$ in XTE~J1701$-$462}
\begin{tabular}{lcccccccc}
\tableline\tableline
HID region & $^{a}R_{\rm in} $ & $^{b}l_{\rm {disk}}$ & $^{c}B_{\rm a}$ & $^{d}\dot{M}_{\rm {disk,\ Edd}}$ & $^{e}l_{\rm {ns}}$ & $^{f}\dot{M}_{\rm {ns,\ Edd}}$ & $^{g}R_{\rm {ns}}$ & $^{h}L_{\rm {total,\ Edd}}$ \\
\tableline 1 (HID1) &
    $12.79_{-0.85}^{+1.04}$ & $1.04_{-0.12}^{+0.15}$    &
    $0.28_{-0.05}^{+0.06}$  & $1.74_{-0.57}^{+0.50}$    &
    $0.47_{-0.10}^{+0.08}$  & $0.123_{-0.027}^{+0.022}$ &
    $8.18_{-1.50}^{+1.33}$  & $0.400_{-0.016}^{+0.016}$ \\
\tableline 2 &
    $20.07_{-1.20}^{+1.54}$ & $1.27_{-0.15}^{+0.20}$    &
    $0.68_{-0.11}^{+0.14}$  & $2.85_{-0.98}^{+0.85}$    &
    $0.47_{-0.10}^{+0.08}$  & $0.123_{-0.026}^{+0.021}$ &
    $8.83_{-1.62}^{+1.45}$  & $0.458_{-0.017}^{+0.019}$ \\
\tableline 3 &
    $20.19_{-1.11}^{+1.40}$ & $1.53_{-0.17}^{+0.21}$    &
    $0.75_{-0.11}^{+0.14}$  & $3.91_{-1.30}^{+1.09}$    &
    $0.49_{-0.08}^{+0.10}$  & $0.130_{-0.021}^{+0.026}$ &
    $9.22_{-0.83}^{+1.01}$  & $0.532_{-0.017}^{+0.022}$ \\
\tableline 4 &
    $21.49_{-1.18}^{+1.52}$ & $1.86_{-0.21}^{+0.26}$    &
    $0.93_{-0.14}^{+0.18}$  & $5.06_{-1.67}^{+1.41}$    &
    $0.55_{-0.10}^{+0.13}$  & $0.144_{-0.027}^{+0.035}$ &
    $9.79_{-1.68}^{+1.96}$  & $0.634_{-0.021}^{+0.027}$ \\
\tableline 5 &
    $22.33_{-1.15}^{+1.45}$ & $2.20_{-0.22}^{+0.27}$     &
    $1.05_{-0.15}^{+0.19}$  & $5.97_{-1.89}^{+1.57}$     &
    $0.60_{-0.11}^{+0.14}$  & $0.158_{-0.029}^{+0.037}$  &
    $10.18_{-1.71}^{+1.88}$ & $0.711_{-0.022}^{+0.029}$ \\
\tableline 6 &
    $22.69_{-1.16}^{+1.49}$ & $2.32_{-0.24}^{+0.30}$    &
    $1.14_{-0.16}^{+0.20}$  & $6.64_{-2.08}^{+1.74}$    &
    $0.64_{-0.12}^{+0.15}$  & $0.168_{-0.032}^{+0.041}$ &
    $10.59_{-1.73}^{+2.11}$ & $0.778_{-0.025}^{+0.032}$ \\
\tableline 7 &
    $23.03_{-1.16}^{+1.47}$ & $2.51_{-0.25}^{+0.32}$    &
    $1.21_{-0.17}^{+0.21}$  & $7.28_{-2.25}^{+1.87}$    &
    $0.68_{-0.13}^{+0.17}$  & $0.178_{-0.033}^{+0.043}$ &
    $11.00_{-1.79}^{+2.11}$ & $0.838_{-0.026}^{+0.034}$ \\
\tableline 8 &
    $23.00_{-1.16}^{+1.52}$ & $2.74_{-0.28}^{+0.35}$    &
    $1.26_{-0.18}^{+0.23}$  & $7.99_{-2.45}^{+2.05}$    &
    $0.70_{-0.14}^{+0.20}$  & $0.184_{-0.037}^{+0.051}$ &
    $10.89_{-1.93}^{+2.37}$ & $0.905_{-0.029}^{+0.038}$ \\
\tableline 9 &
    $23.94_{-1.04}^{+1.50}$ & $3.10_{-0.27}^{+0.39}$    &
    $1.44_{-0.17}^{+0.25}$  & $9.50_{-2.71}^{+2.41}$    &
    $0.68_{-0.12}^{+0.24}$  & $0.180_{-0.032}^{+0.064}$ &
    $10.67_{-2.06}^{+2.54}$ & $0.996_{-0.027}^{+0.044}$ \\
\tableline 10 &
    $24.95_{-1.11}^{+1.49}$ & $3.22_{-0.29}^{+0.38}$    &
    $1.58_{-0.19}^{+0.26}$  & $10.16_{-2.93}^{+2.47}$   &
    $0.60_{-0.14}^{+0.21}$  & $0.159_{-0.036}^{+0.055}$ &
    $10.67_{-2.14}^{+2.68}$ & $1.005_{-0.029}^{+0.041}$ \\
\tableline 11 &
    $25.97_{-0.97}^{+1.11}$ & $3.23_{-0.24}^{+0.28}$    &
    $1.70_{-0.18}^{+0.20}$  & $11.62_{-2.88}^{+2.07}$   &
    $0.48_{-0.08}^{+0.09}$  & $0.126_{-0.022}^{+0.025}$ &
    $9.39_{-1.23}^{+1.42}$  & $0.975_{-0.023}^{+0.026}$ \\
\tableline 12 &
    $25.97_{-0.84}^{+0.94}$ & $3.18_{-0.20}^{+0.23}$    &
    $1.68_{-0.15}^{+0.17}$  & $10.37_{-2.64}^{+1.80}$   &
    $0.34_{-0.06}^{+0.07}$  & $0.090_{-0.017}^{+0.019}$ &
    $7.63_{-1.09}^{+1.20}$  & $0.927_{-0.018}^{+0.021}$ \\
\tableline 13 &
    $28.00_{-1.21}^{+1.81}$ & $3.04_{-0.26}^{+0.40}$    &
    $1.87_{-0.22}^{+0.34}$  & $10.83_{-3.21}^{+2.87}$   &
    $0.31_{-0.10}^{+0.20}$  & $0.082_{-0.027}^{+0.052}$ &
    $8.80_{-2.62}^{+3.85}$  & $0.883_{-0.025}^{+0.041}$ \\
\tableline 14 &
    $26.81_{-0.65}^{+0.74}$ & $3.00_{-0.15}^{+0.16}$    &
    $1.73_{-0.12}^{+0.13}$  & $10.18_{-2.40}^{+1.53}$   &
    $0.16_{-0.04}^{+0.05}$  & $0.042_{-0.010}^{+0.012}$ &
    $5.18_{-0.89}^{+1.07}$  & $0.832_{-0.013}^{+0.015}$ \\
\tableline 15 &
    $27.38_{-0.65}^{+0.74}$ & $2.87_{-0.14}^{+0.16}$    &
    $1.75_{-0.11}^{+0.13}$  & $9.92_{-2.37}^{+1.51}$    &
    $0.10_{-0.03}^{+0.04}$  & $0.028_{-0.008}^{+0.011}$ &
    $4.07_{-0.99}^{+1.26}$  & $0.783_{-0.011}^{+0.014}$ \\
\tableline 16 &
    $26.69_{-0.89}^{+1.06}$ & $2.61_{-0.18}^{+0.21}$    &
    $1.60_{-0.15}^{+0.17}$  & $8.86_{-2.39}^{+1.68}$    &
    $0.06_{-0.03}^{+0.04}$  & $0.016_{-0.007}^{+0.011}$ &
    $2.71_{-0.93}^{+1.41}$  & $0.704_{-0.014}^{+0.017}$ \\
\tableline 17 &
    $25.75_{-1.11}^{+0.87}$ & $2.39_{-0.21}^{+0.16}$    &
    $1.44_{-0.17}^{+0.13}$  & $7.78_{-2.33}^{+1.32}$    &
    $0.03_{-0.01}^{+0.01}$  & $0.009_{-0.002}^{+0.004}$ &
    $1.60_{-0.29}^{+0.39}$  & $0.639_{-0.015}^{+0.012}$ \\
\tableline 18 &
    $24.00_{-0.84}^{+1.13}$ & $1.72_{-0.07}^{+0.10}$    &
    $1.08_{-0.09}^{+0.12}$  & $5.33_{-1.29}^{+0.86}$    &
    $0.02_{-0.01}^{+0.01}$  & $0.005_{-0.002}^{+0.002}$ &
    $0.91_{-0.19}^{+0.21}$  & $0.458_{-0.005}^{+0.008}$ \\
\tableline\tableline 1 (HID2) &
    $14.36_{-0.99}^{+1.23}$ & $0.87_{-0.12}^{+0.15}$    &
    $0.31_{-0.06}^{+0.07}$  & $0.33_{-0.07}^{+0.08}$    &
    $0.44_{-0.07}^{+0.08}$  & $0.116_{-0.018}^{+0.021}$ &
    $11.89_{-1.81}^{+1.98}$ & $0.346_{-0.013}^{+0.016}$ \\
\tableline 2 &
    $15.34_{-1.03}^{+1.33}$ & $0.96_{-0.13}^{+0.17}$    &
    $0.37_{-0.07}^{+0.09}$  & $1.58_{-0.59}^{+0.51}$    &
    $0.44_{-0.08}^{+0.09}$  & $0.116_{-0.020}^{+0.023}$ &
    $12.59_{-2.04}^{+2.33}$ & $0.369_{-0.014}^{+0.018}$ \\
\tableline 3 &
    $15.00_{-0.77}^{+0.87}$ & $1.26_{-0.13}^{+0.15}$    &
    $0.41_{-0.06}^{+0.07}$  & $2.40_{-0.73}^{+0.57}$    &
    $0.38_{-0.05}^{+0.06}$  & $0.100_{-0.014}^{+0.015}$ &
    $10.32_{-1.14}^{+1.18}$ & $0.431_{-0.013}^{+0.014}$  \\
\tableline 4 &
    $15.20_{-0.68}^{+0.77}$ & $1.29_{-0.12}^{+0.13}$    &
    $0.42_{-0.05}^{+0.06}$  & $2.48_{-0.71}^{+0.53}$    &
    $0.32_{-0.05}^{+0.05}$  & $0.083_{-0.012}^{+0.013}$ &
    $9.45_{-1.05}^{+1.18}$  & $0.422_{-0.011}^{+0.012}$ \\
\tableline 5 &
    $14.96_{-0.60}^{+0.68}$ & $1.40_{-0.11}^{+0.13}$    &
    $0.43_{-0.05}^{+0.05}$  & $2.63_{-0.70}^{+0.52}$    &
    $0.22_{-0.04}^{+0.05}$  & $0.058_{-0.012}^{+0.013}$ &
    $8.12_{-1.28}^{+1.47}$  & $0.425_{-0.011}^{+0.012}$ \\
\tableline 6 &
    $17.95_{-0.94}^{+1.13}$ & $1.50_{-0.16}^{+0.19}$    &
    $0.61_{-0.09}^{+0.10}$  & $3.40_{-1.08}^{+0.86}$    &
    $0.19_{-0.05}^{+0.06}$  & $0.051_{-0.012}^{+0.015}$ &
    $7.63_{-1.47}^{+1.73}$  & $0.445_{-0.014}^{+0.017}$ \\
\tableline 7 &
    $11.61_{-0.43}^{+0.51}$ & $1.49_{-0.11}^{+0.13}$    &
    $0.28_{-0.03}^{+0.03}$  & $2.18_{-0.52}^{+0.41}$    &
    $0.15_{-0.05}^{+0.07}$  & $0.041_{-0.014}^{+0.019}$ &
    $5.34_{-1.43}^{+1.88}$  & $0.430_{-0.011}^{+0.014}$ \\
\tableline 8 &
    $9.40_{-0.24}^{+0.29}$  & $1.57_{-0.08}^{+0.10}$    &
    $0.20_{-0.02}^{+0.02}$  & $1.87_{-0.36}^{+0.26}$    &
    $0.10_{-0.04}^{+0.06}$  & $0.026_{-0.010}^{+0.017}$ &
    $2.80_{-0.99}^{+1.45}$  & $0.439_{-0.008}^{+0.011}$ \\
\tableline 9 &
    $8.14_{-0.21}^{+0.27}$  & $1.59_{-0.08}^{+0.11}$    &
    $0.16_{-0.02}^{+0.02}$  & $1.63_{-0.29}^{+0.23}$    &
    $0.09_{-0.04}^{+0.08}$  & $0.024_{-0.011}^{+0.022}$ &
    $2.40_{-1.04}^{+1.70}$  & $0.442_{-0.008}^{+0.013}$ \\
\tableline 10 &
    $6.94_{-0.27}^{+0.51}$  & $1.73_{-0.13}^{+0.25}$    &
    $0.12_{-0.02}^{+0.03}$  & $1.52_{-0.32}^{+0.39}$    &
    $0.12_{-0.08}^{+0.06}$  & $0.031_{-0.021}^{+0.017}$ &
    $2.86_{-1.76}^{+1.93}$  & $0.487_{-0.015}^{+0.022}$ \\
\tableline 11 &
    $5.71_{-0.15}^{+0.39}$  & $1.98_{-0.10}^{+0.27}$    &
    $0.09_{-0.01}^{+0.02}$  & $1.43_{-0.23}^{+0.34}$    &
    $0.03_{-0.06}^{+0.05}$  & $0.008_{-0.015}^{+0.013}$ &
    $0.88_{-1.33}^{+1.66}$  & $0.528_{-0.011}^{+0.022}$ \\
\tableline
\end{tabular}
\tablenotetext{a}{The inner disk radius for an inclination angle ($\theta$) of $70^\circ$, in unit of km}
\tablenotetext{b}{The unabsorbed MCD luminosity for a distance of 8.8 kpc to the source, in units of $10^{38}$~ergs~s$^{-1}$}
\tablenotetext{c}{The NS apparent magnetic field strength, in units of $10^9$~G}
\tablenotetext{d}{The disk accretion rate ($\dot{M}_{\rm {disk}}$), in unit of $\dot{M}_{\rm Edd}$}
\tablenotetext{e}{The unabsorbed NS luminosity for a distance of 8.8 kpc to the source, in units of $10^{38}$~ergs~s$^{-1}$}
\tablenotetext{f}{The NS accretion rate ($\dot{M}_{\rm {ns}}$), in unit of $\dot{M}_{\rm Edd}$}
\tablenotetext{g}{The NS radius derived from the BB temperature and luminosity and
multiplied by the square of a factor $\sim$1.6 due to color correction, in unit of km}
\tablenotetext{h}{The total unabsorbed
luminosity ($L_{\rm total}$), in unit of $L_{\rm Edd}$}
\end{table}

\clearpage

\begin{figure}[t]
\centerline{
\includegraphics[width=6.0cm,height=12.5cm,angle=-90]{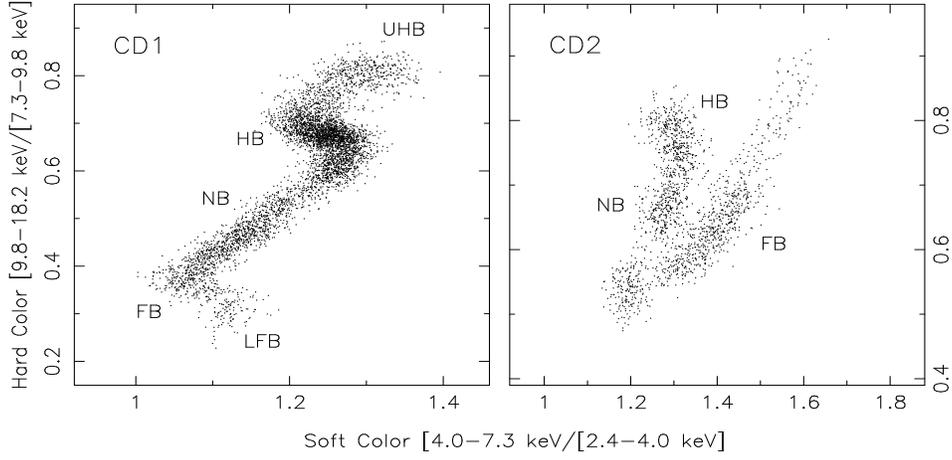}}
\vspace{0.25cm}
\caption{Two CDs of XTE~J1701$-$462, which are defined in \S 2. Each
point represents 16 s background-subtracted data from PCU2. CD1
includes a ``Z'' track and the track of CD2 resembles the character ``$\nu$''.}
\end{figure}

\begin{figure}[t]
\centerline{
\includegraphics[width=6.0cm,height=12.5cm,angle=-90]{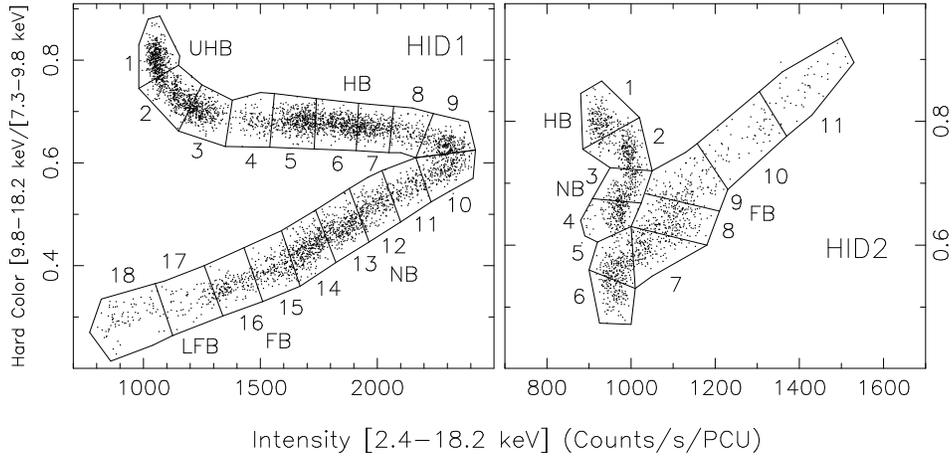}}
\vspace{0.25cm}
\caption{Two HIDs of XTE~J1701$-$462, which are defined in \S 2. Each
point represents 16 s background-subtracted data from PCU2. HID1 and
HID2 are separately divided into eighteen regions (labeled 1,2,3,...,18)
and eleven regions (labeled 1, 2, 3,...,11), in order to group data and
extract the PCA+HEXTE spectrum of each region of the 29 regions.}
\end{figure}

\clearpage

\begin{figure*}[t]
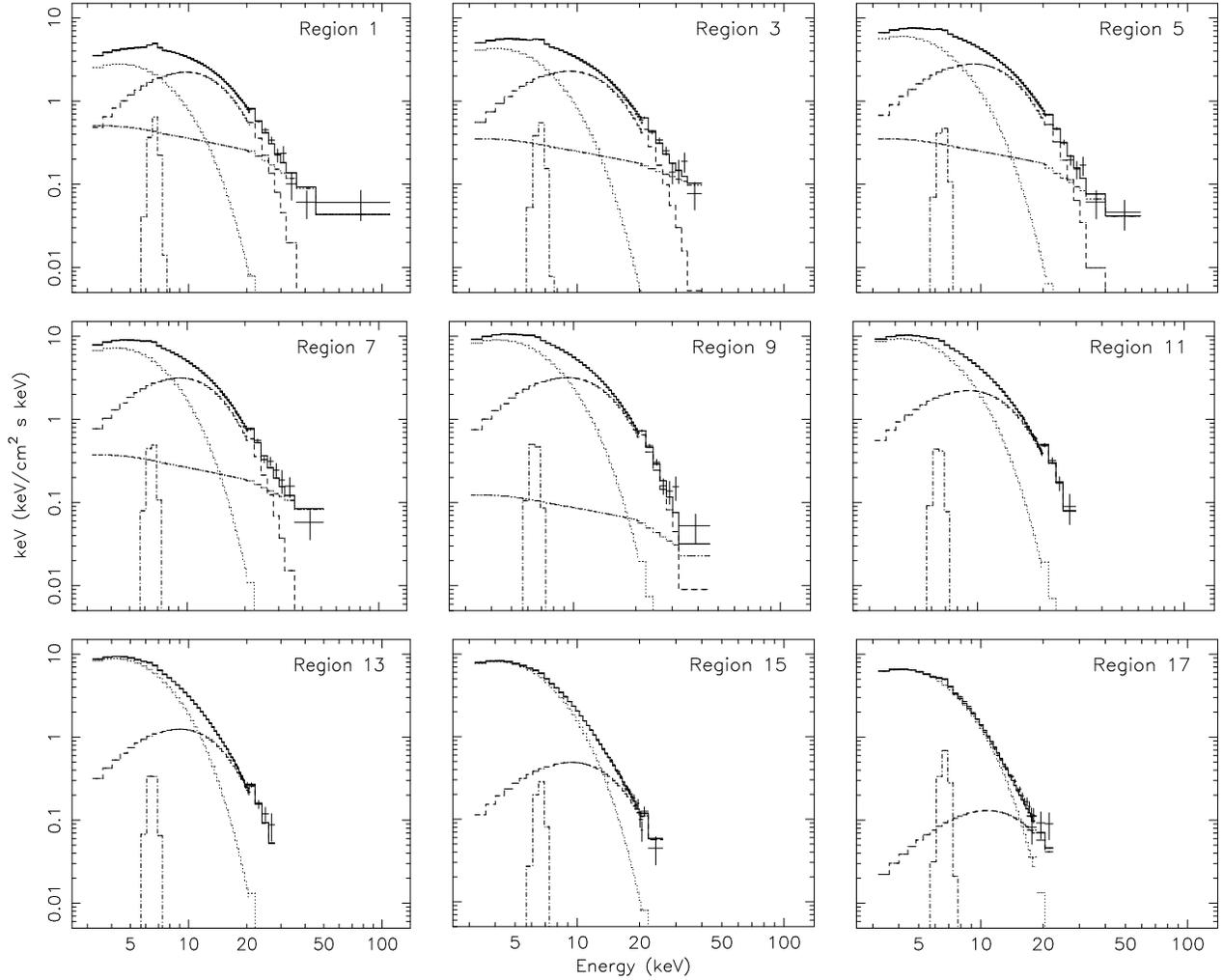

\centerline{\hbox{
\includegraphics[width=4.cm,height=5.0cm,angle=-90]{f3_1.ps}
\hspace{0.35cm}
\includegraphics[width=4.cm,height=5.0cm,angle=-90]{f3_2.ps}
\hspace{0.35cm}
\includegraphics[width=4.cm,height=5.0cm,angle=-90]{f3_3.ps}}}
\vspace{0.35cm} \centerline{\hspace{-0.6cm}{
\includegraphics[width=4.cm,height=5.5cm,angle=-90]{f3_4.ps}
\hspace{0.35cm}
\includegraphics[width=4.cm,height=5.0cm,angle=-90]{f3_5.ps}
\hspace{0.35cm}
\includegraphics[width=4.cm,height=5.0cm,angle=-90]{f3_6.ps}}}
\vspace{0.35cm} \centerline{\hbox{
\includegraphics[width=4.3cm, height=5.cm,angle=-90]{f3_7.ps}
\hspace{0.35cm}
\includegraphics[width=4.65cm,height=5.0cm,angle=-90]{f3_8.ps}
\hspace{0.35cm}
\includegraphics[width=4.3cm,height=5.0cm,angle=-90]{f3_9.ps}}}
\vspace{0.25cm}
\caption{Nine unfolded spectra of HID1 regions, showing four
components, namely, MCD (dashed line), LINE (dot-dashed line),
BB (dotted line), and CBPL (dot-dot-dashed line), and the
sum of the individual components (solid line) for the UHB
and HB spectra, while only three components, i.e., MCD, BB,
and LINE, for the NB, FB, and LFB spectra.}
\end{figure*}

\clearpage

\begin{figure*}[t]
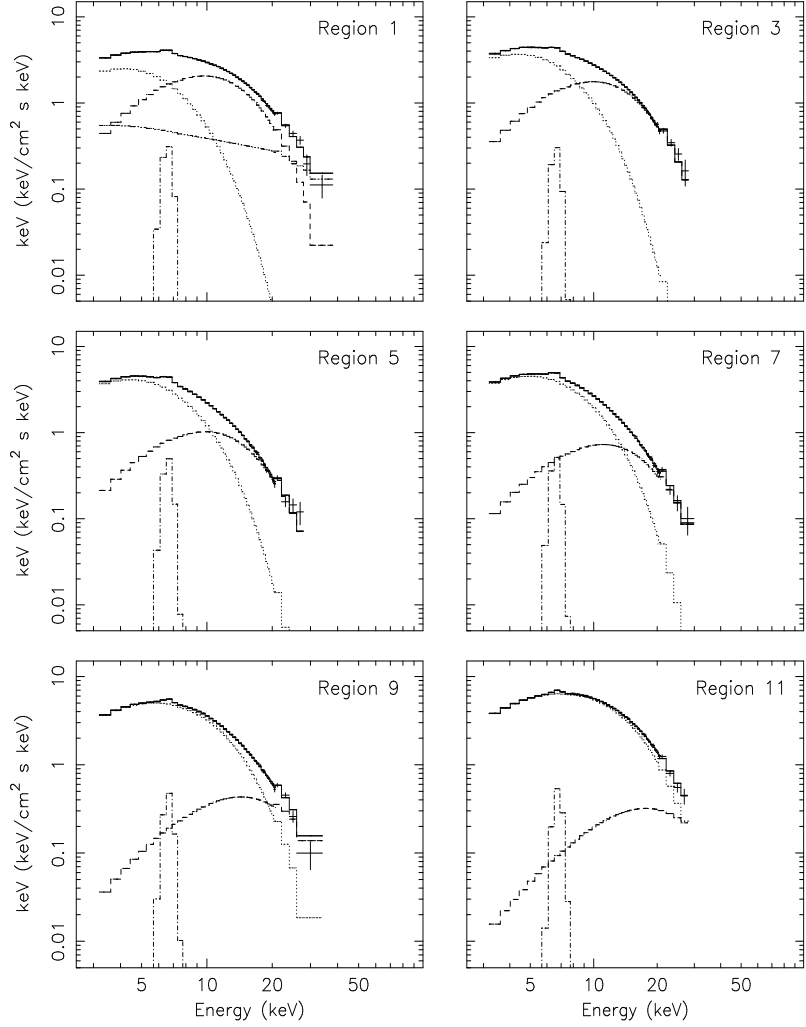

\centerline{\hbox{
\includegraphics[width=4.cm,height=5.5cm,angle=-90]{f4_1.ps}
\hspace{0.35cm}
\includegraphics[width=4.cm,height=4.5cm,angle=-90]{f4_2.ps}}}
\vspace{0.35cm} \centerline{\hbox{
\includegraphics[width=4.cm,height=5.5cm,angle=-90]{f4_3.ps}
\hspace{0.35cm}
\includegraphics[width=4.cm,height=4.5cm,angle=-90]{f4_4.ps}}}
\vspace{0.35cm} \centerline{\hbox{
\includegraphics[width=4.8cm,height=5.5cm,angle=-90]{f4_5.ps}
\hspace{0.35cm}
\includegraphics[width=4.8cm,height=4.5cm,angle=-90]{f4_6.ps}}}
\vspace{0.25cm}
\caption{Six unfolded spectra of HID2 regions, showing four
components [MCD (dashed line), LINE (dot-dashed line), BB
(dotted line), and CBPL (dot-dot-dashed line)] and the
sum of the individual components (solid line) for the HB
spectra, while only three components (MCD, BB, and LINE)
for the NB and FB spectra.}
\end{figure*}

\clearpage

\begin{figure*}[t]
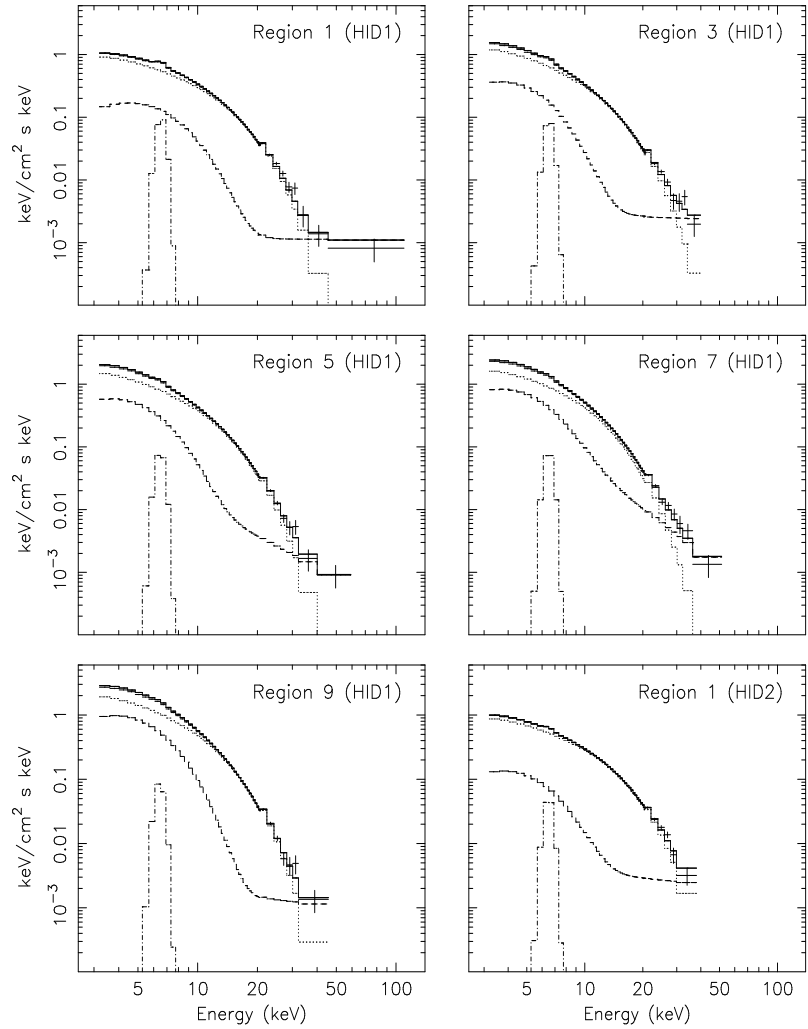

\centerline{\hbox{
\includegraphics[width=4.cm,height=5.5cm,angle=-90]{f5_1.ps}
\hspace{0.35cm}
\includegraphics[width=4.cm,height=4.5cm,angle=-90]{f5_2.ps}}}
\vspace{0.35cm} \centerline{\hbox{
\includegraphics[width=4.cm,height=5.5cm,angle=-90]{f5_3.ps}
\hspace{0.35cm}
\includegraphics[width=4.cm,height=4.5cm,angle=-90]{f5_4.ps}}}
\vspace{0.35cm} \centerline{\hbox{
\includegraphics[width=4.8cm,height=5.5cm,angle=-90]{f5_5.ps}
\hspace{0.35cm}
\includegraphics[width=4.8cm,height=4.5cm,angle=-90]{f5_6.ps}}}
\vspace{0.25cm}
\caption{Six unfolded spectra of the UHB and HBs, showing three
components, i.e., BMC (dashed line), CompTT (dotted line), and
LINE (dot-dashed line), and the sum (solid line) of the individual
components. The crosses represent the data.}
\end{figure*}

\clearpage

\begin{figure*}[t]
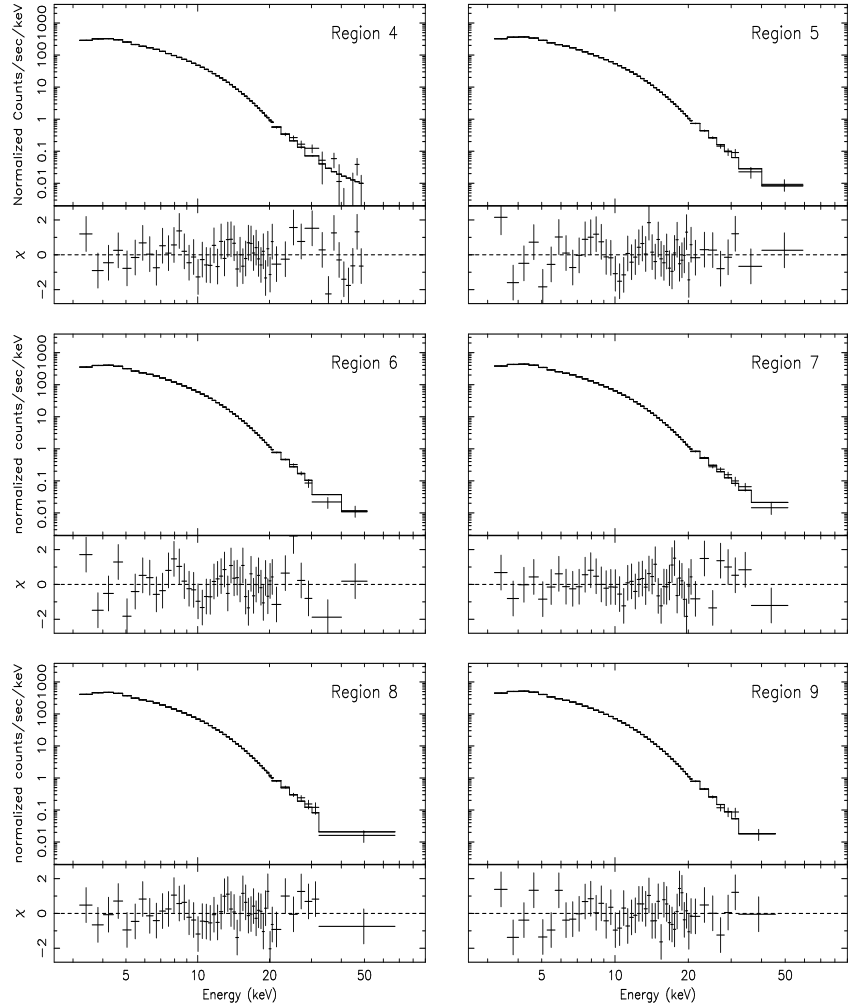

\centerline{\hbox{
\includegraphics[width=4.cm,height=5.5cm,angle=-90]{f6_1.ps}
\hspace{0.35cm}
\includegraphics[width=4.cm,height=5.05cm,angle=-90]{f6_2.ps}}}
\vspace{0.35cm} \centerline{\hbox{
\includegraphics[width=4.cm,height=5.5cm,angle=-90]{f6_3.ps}
\hspace{0.35cm}
\includegraphics[width=4.cm,height=5.05cm,angle=-90]{f6_4.ps}}}
\vspace{0.35cm} \centerline{\hbox{
\includegraphics[width=4.53cm,height=5.5cm,angle=-90]{f6_5.ps}
\hspace{0.35cm}
\includegraphics[width=4.53cm,height=5.05cm,angle=-90]{f6_6.ps}}}
\vspace{0.25cm}
\caption{The data, folded spectra, and the corresponding residuals of the HB spectra of HID1, usin the MCD+BB+CPBL+LINE model.}
\end{figure*}

\clearpage

\begin{figure*}[t]
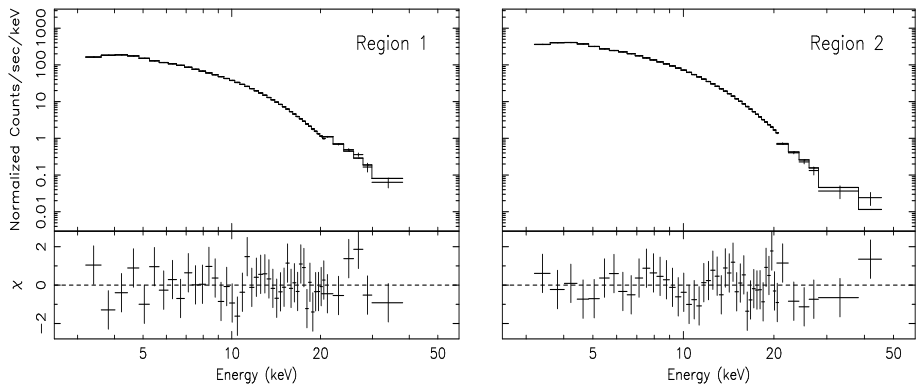

\centerline{\hbox{
\includegraphics[width=5.cm,height=6.cm,angle=-90]{f7_1.ps}
\hspace{0.35cm}
\includegraphics[width=5.cm,height=5.5cm,angle=-90]{f7_2.ps}}}
\vspace{0.25cm}
\caption{The data, folded spectra, and the corresponding residuals of the HB spectra of HID2, usin the BMC+CompTT+LINE model.}
\end{figure*}

\clearpage

\begin{figure*}[t]
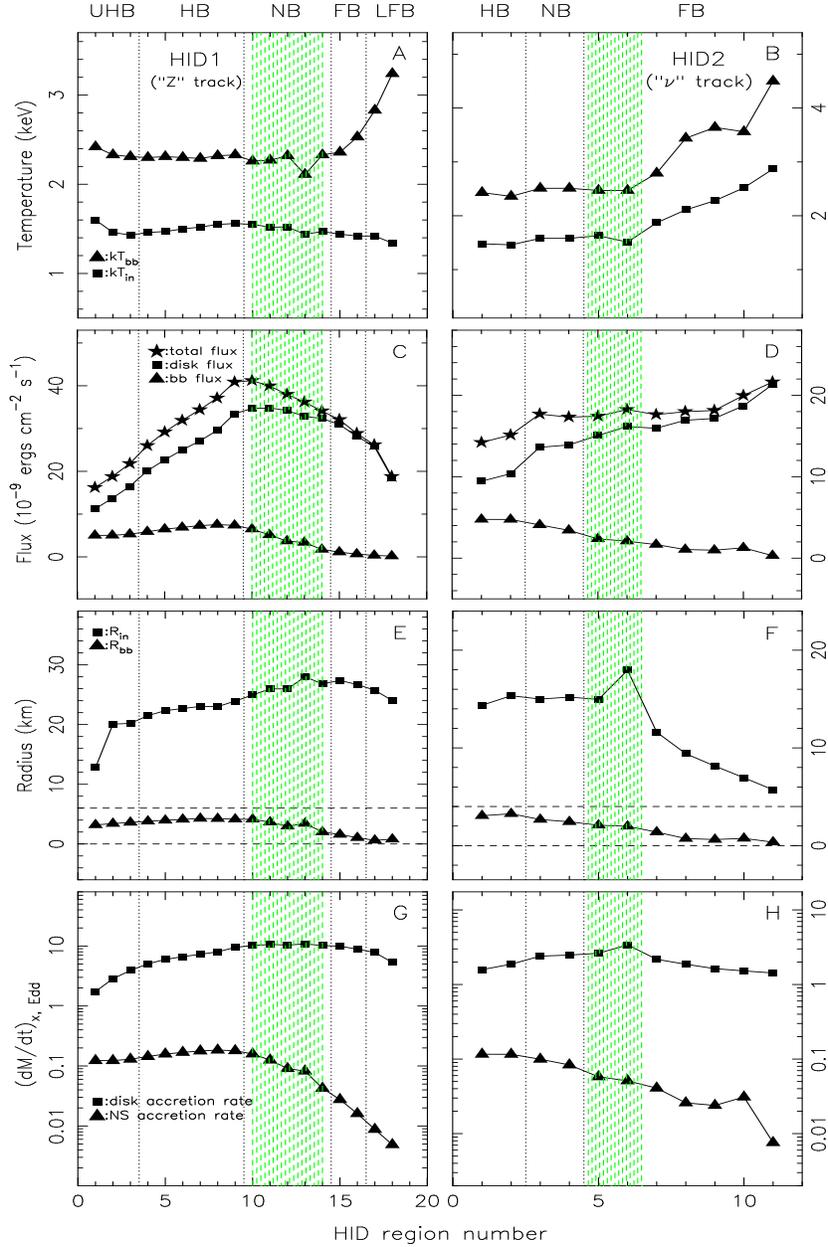

\centerline{\hbox{
\includegraphics[width=4.2cm,height=10.8cm,angle=-90]{f8_1.ps}}}
\vspace{0.1cm}
\centerline{\hspace{-0.07cm}{
\includegraphics[width=3.6cm,height=10.86cm,angle=-90]{f8_2.ps}}}
\vspace{0.1cm} \centerline{\hbox{
\includegraphics[width=3.6cm,height=10.8cm,angle=-90]{f8_3.ps}}}
\vspace{0.1cm} \centerline{\hbox{
\includegraphics[width=4.7cm,height=10.8cm,angle=-90]{f8_4.ps}}}
\vspace{0.25cm}
\caption{The evolutions of the fitting parameters and mass accretion
rates along the HID tracks. The results are obtained from the PCA+HEXTE
spectral fitting, using the MCD+BB+CBPL+LINE model for the UHB and HB
spectra, and the MCD+BB+LINE model for the NB, FB, and LFB spectra. The
squares and triangles represent the results obtained from the MCD and BB,
respectively. The green regions are the transition ones.}
\end{figure*}

\clearpage

\begin{figure}[t]
\centerline{
\includegraphics[width=6.5cm,height=8.cm,angle=-90]{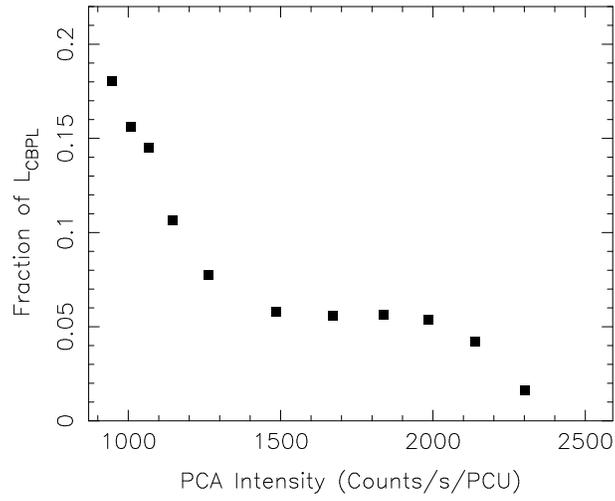}}
\vspace{0.25cm}
\caption{Fraction of the $L_{\rm CBPL}$ in the Z-stages, including a
Cyg-like stage (the ``Z'' track) and a Sco-like stage (the ``$\nu$'' track).
The luminosities are estimated in the 1.5--200 keV interval.}
\end{figure}

\clearpage

\begin{figure*}[t]
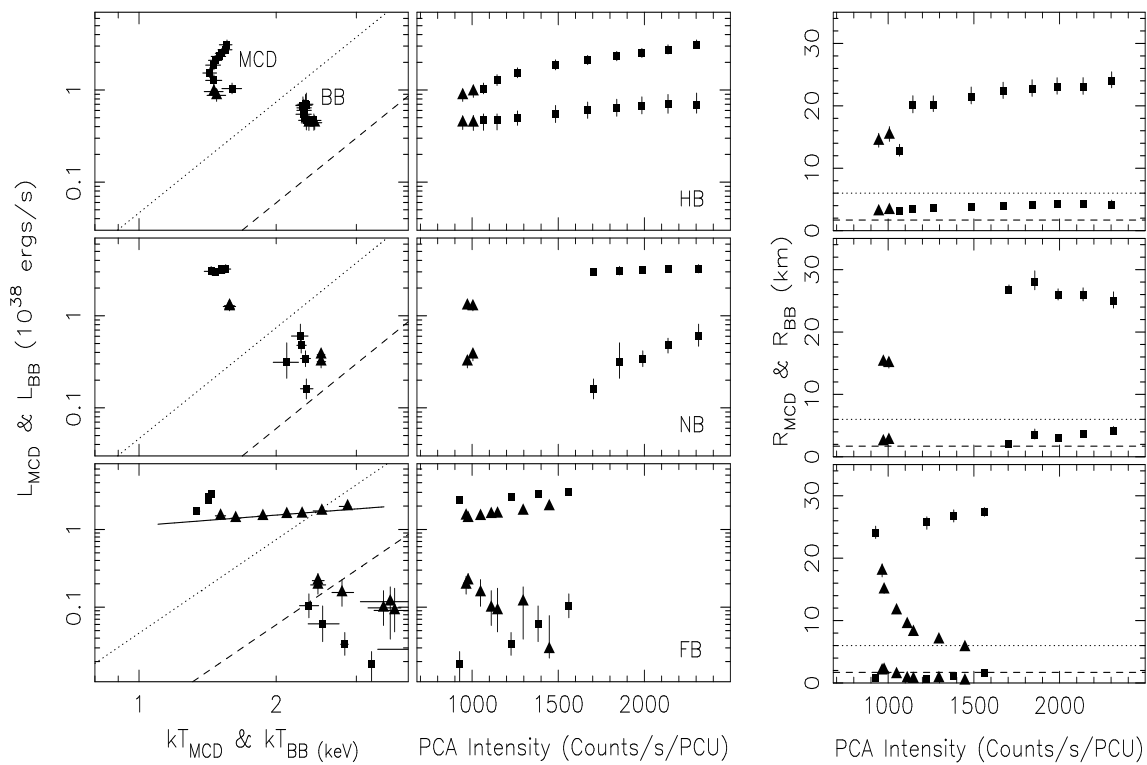

\centerline{\hbox{
\includegraphics[width=10.cm,height=9.6cm,angle=-90]{f10_1.ps}
\hspace{0.3cm}
\includegraphics[width=10.cm,height=5.cm,angle=-90]{f10_2.ps}}}
\vspace{0.25cm}
\caption{Spectral fitting results for a ``Z'' track and a ``$\nu$'' track.
The luminosities of thermal components (MCD/BB) vs. their characteristic
temperatures and as well as the PCA intensity are shown by the left column and
mid column, respectively. The right column shows the characteristic emission
sizes of the thermal components vs. the PCA intensity. The top, mid, and
bottom rows correspond to the UHB and HB interval, NB interval, and FB
and LFB interval, respectively. The squares and triangles separately
represent the results obtained from HID1 and HID2. The dotted line and
dashed line correspond to $R$ = 6.0 cm and $R$ = 1.7 km, respectively;
in the left column, assuming $L_{\rm X}=4\pi R^{2}{\sigma}_{\rm SB}T^{4}$.
In the bottom panel of the left column, the solid line corresponds
to $L_{\rm X}\propto T^{0.45}$. In each panel, the results of the MCD are
always at the top, and those of the BB are at the
bottom. $kT_{\rm MCD}$ = $kT_{\rm in}$, $L_{\rm MCD}$ = $L_{\rm disk}$, $L_{\rm BB}$ = $L_{\rm ns}$.}
\end{figure*}

\clearpage

\begin{figure*}[t]
\centerline{
\includegraphics[width=12.cm,height=12.cm,angle=-90]{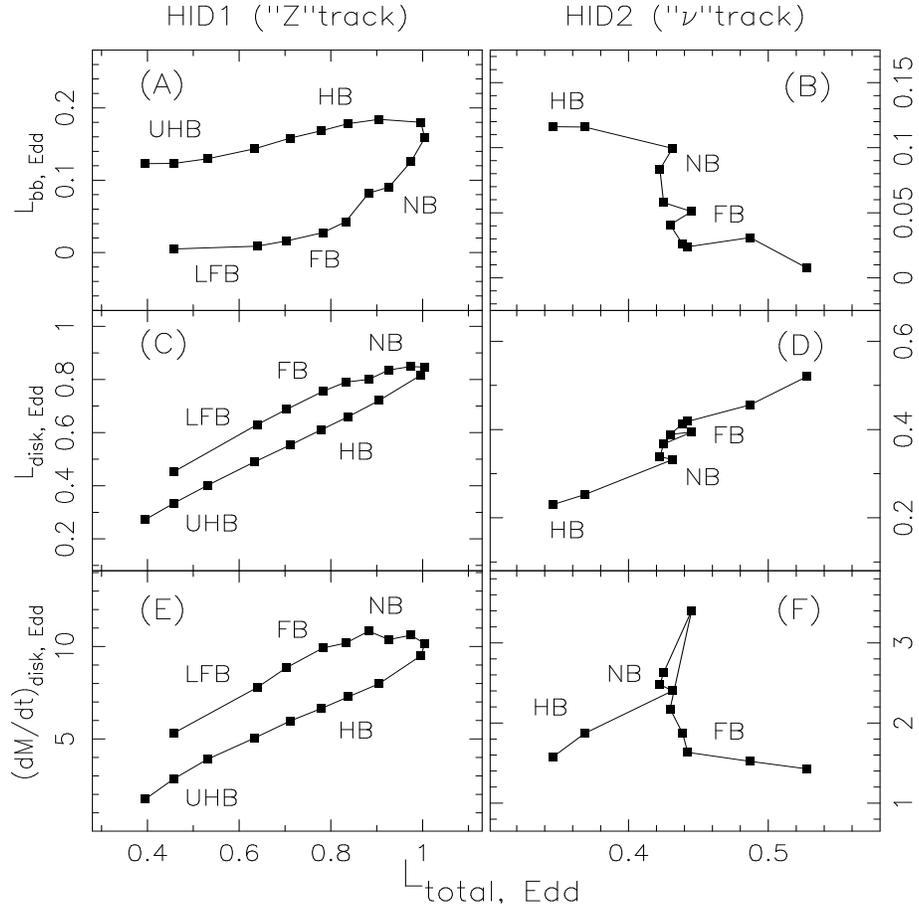}}
\vspace{0.25cm}
\caption{Correlations between the total luminosity and the luminosities
of the thermal components as well as the disk mass accration rate. The left
and right panels represent the results obtained from HID1 and HID2, respectively.}
\end{figure*}

\clearpage

\begin{figure}[t]
\centerline{
\includegraphics[width=8.cm,height=13.8cm,angle=-90]{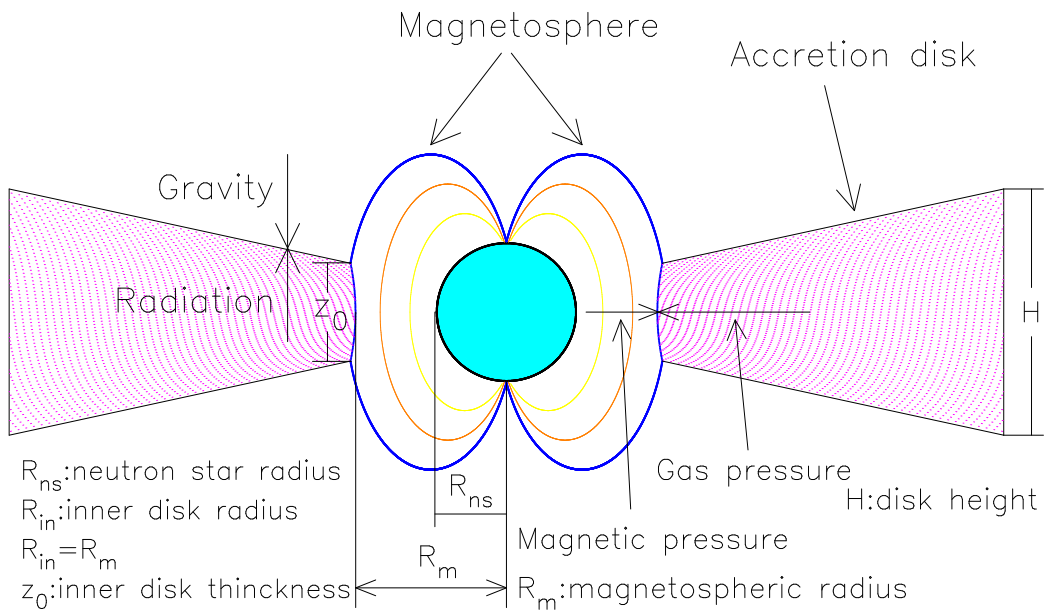}}
\vspace{0.25cm}
\caption{Schematic view of the proposed geometry for a NSXB hosting a
NS, a magnetosphere, and a radiation pressure dominated accretion disk.
The inner disk radius is defined as the radius at which the magnetic
pressure in the magnetosphere balances the gas pressure in the accretion
disk. The vertical thickness of the inner disk region is determined by
the balance between the vertical gravity of the NS and the radiation
pressure in the disk.}
\end{figure}

\clearpage

\begin{figure*}[t]
\centerline{
\includegraphics[width=10.5cm,height=14.cm,angle=-90]{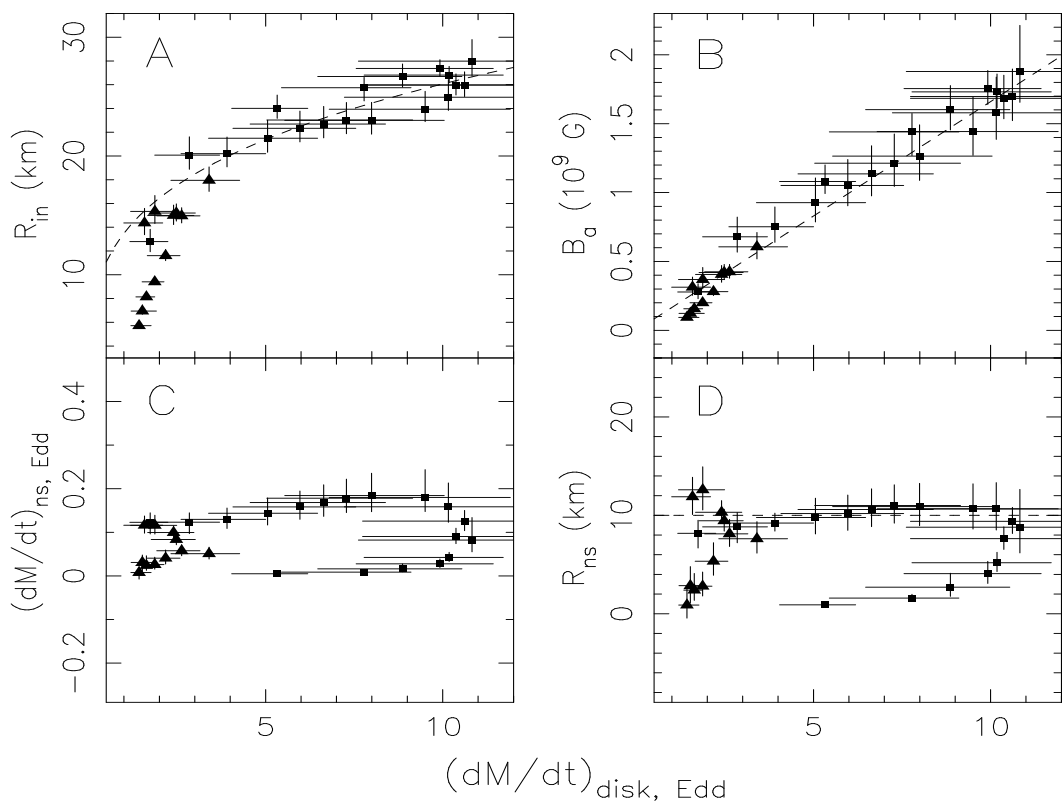}}
\vspace{0.25cm}
\caption{Correlations of the disk accretion rate with several
inferred parameters of the accretion disk and the NS. The squares and triangles
represent the results obtained from HID1 and HID2, respectively, assuming the
inclination angle to be $70^\circ$. (A) The correlation between the disk accretion
rate and the inner disk radius. The reference line corresponds
to $R_{\rm in}=14\dot{M}_{\rm disk}^{2/7}$. (B) The disk accretion rate vs. the inferred
NS apparent magnetic field strength. The reference line corresponds to $B=0.17\dot{M}_{\rm disk}$.
(C) The correlation between the disk accretion rate and the NS accretion rate. (D)
The disk accretion rate vs. the inferred NS radius. The reference line corresponds
to $R_{\rm ns}=10$ km.}
\end{figure*}

\clearpage

\begin{figure}[t]
\centerline{
\includegraphics[width=18.0cm]{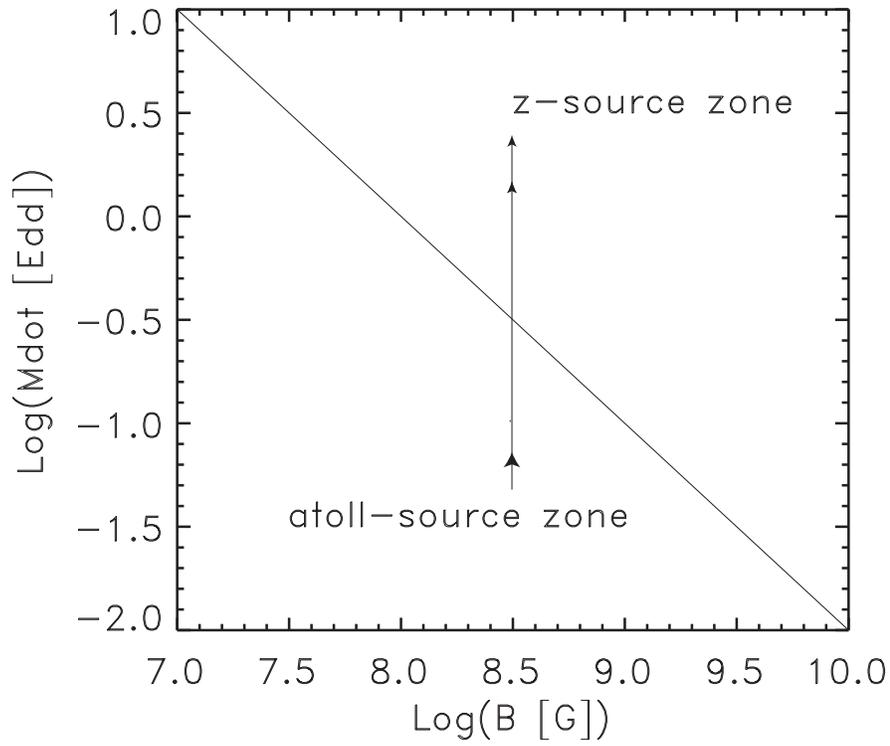}}
\vspace{-1.4cm}
\caption{Illustration of the proposed atoll and Z source zones. The parameter
space above the diagonal line is the Z-source zone and the region below the
line is the atoll-source zone. Therefore if a source increases its accretion
rate, i.e., moves vertically in the plot, it may cross the line and changes
its behaviors from an atoll source to a Z-source, or vice versa, such as the case
for XTE~J1701$-$462.}
\end{figure}

\end{document}